\documentclass[twocolumn,english,pra, eqsecnum]{revtex4-1}
\usepackage[T1]{fontenc}
\usepackage[latin9]{inputenc}
\usepackage{color}
\usepackage{bm}
\usepackage{amsmath}
\usepackage{amssymb}
\usepackage{stackrel}
\usepackage{graphicx}
\usepackage{wasysym}
\usepackage{esint}

\makeatletter

\providecommand{\tabularnewline}{\\}

\usepackage{braket}
\date{}
\usepackage{caption}
\makeatother

\usepackage{babel}
\begin{document}
\title{The fate of the false vacuum: Finite temperature, entropy and topological
phase in quantum simulations of the early universe}
\author{King Lun Ng, Bogdan Opanchuk, Manushan Thenabadu, Margaret Reid and
Peter D. Drummond}
\affiliation{Centre for Quantum and Optical Science, Swinburne University of Technology,
Melbourne 3122, Australia}
\begin{abstract}
Despite being at the heart of the theory of the ``Big Bang'' and
cosmic inflation, the quantum field theory prediction of false vacuum
tunneling has not been tested. To address the exponential complexity of the problem, a table-top quantum simulator in the form of an engineered Bose-Einstein condensate (BEC) has been proposed to give dynamical solutions of the quantum field equations. In this paper, we give a numerical feasibility study of the BEC quantum simulator under realistic conditions
and temperatures, with an approximate truncated Wigner (tW) phase-space
method. We report the observation of false vacuum tunneling in these
simulations, and the formation of multiple bubble 'universes' with
distinct topological properties. The tunneling gives a transition
of the relative phase of coupled Bose fields from a metastable to
a stable 'vacuum'. We include finite temperature effects that would
be found in a laboratory experiment and also analyze the cut-off dependence
of modulational instabilities in Floquet space. Our numerical phase-space
model does not use thin-wall approximations, which are inapplicable to cosmologically interesting models. It is expected to give the correct quantum treatment including superpositions and entanglement during dynamics.
By analyzing a nonlocal observable called the topological phase entropy
(TPE), our simulations provide information about phase structure in
the true vacuum. We observe a cooperative effect in which the true
vacua bubbles representing distinct universes each have one or the
other of two distinct topologies. The TPE initially increases with
time, reaching a peak as the multiple universes are formed, and then
decreases with time to the phase-ordered vacuum state. This gives
a model for the formation of universes with one of two distinct phases, which is  a possible solution to the problem of particle-antiparticle asymmetry.
\end{abstract}
\maketitle

\section{Introduction}

The evolution of the early universe described by inflation is now
a standard model of cosmic evolution. This theory is widely accepted
because of the observational evidence including the cosmic microwave
background radiation (CMB) detected in all directions. This evidence
points to a ``Big Bang'' origin as the beginning of the Universe.
One of the building blocks is a quantum field theory (QFT) model that
explains the origin of the early universe and the observed temperature
fluctuations in the CMB. These effects originated in Coleman's theory
of quantum tunneling of a scalar quantum field in an initially metastable
vacuum \citep{Coleman1,Coleman2}. The validity of the thin-wall approximations
used in this theory and later variations are not verified. Quantum
field models are exponentially complex and impossible to solve directly. 

It is clearly not possible to repeat the start of the universe, so
it has been proposed to verify the solutions experimentally. Such
experiments would effectively be an analog quantum computer for the
scalar field dynamics of the universe. Here we note that the geometry
should be multi-mode and have no boundaries, to allow free-space nucleation.
The proposal for a suitable quantum analog computer uses a two-species
Bose-Einstein condensate (BEC) experiment in a one-dimensional uniform
ring configuration, similar to those studied in several laboratories
\citep{guo2020supersonic,hopkins2004proposed,bell2016bose}. The Bose-Einstein
condensate has a modulated coupling between two spin components, which
creates a local minimum in the effective phase potential. This allows
an experimental study of models of false vacuum quantum tunneling
in relativistic scalar field theories, with an engineered scalar field
potential.

It is essential to also model the quantum simulator itself. Firstly,
this provides insight into the quantum equations themselves, even
if only in an approximation. Secondly, it is necessary to understand
the performance of the BEC quantum simulator, and its realization
in the laboratory as far as possible. In this paper, we employ numerical
phase-space methods to simulate the quantum dynamics of the BEC quantum
simulator under the conditions of laboratory experiments for up to
$1024$ modes. Earlier analyses have verified a false vacuum tunneling,
but have not accounted for the effects of finite temperatures and
Floquet instabilities in such experiments. Our results show that a
momentum cutoff is essential, and that a low initial temperature is
required.

A feature of this scalar field model is the existence of a spontaneously
broken discrete phase symmetry. This leads to nonlocal topological
effects, which are uncovered using phase-unwrapping image processing
analysis on the data. There are two distinct types of quantum vacua
created with opposite phases. These are locally identical but globally
distinct. Similar models have been employed as possible explanations
of particle-antiparticle asymmetry \citep{Sakharov_1991,zel1974cosmological,kibble1980some}.
We show that such global topological effects can be quantified using
the concept of an observational phase entropy, which can both increase
and decrease with time.

Each true vacuum created from the decay of metastable vacua can expand
into a separate ``universe'' with different topological phases,
creating domains of vacuum with boundaries. The result of nucleation
includes the tunneling rate, the fluctuations in density and temperature,
and the collision of domain walls. All play an important role in the
physical nature of the resulting universe formed. However, theoretical
work so far relies on a number of assumptions which have not been
tested in an experiment. False vacuum decay in zero space dimensions
has been recently simulated using a quantum computer \citep{Spannowsky2020},
but this uses many orders of magnitude fewer qubits than are needed
in a spatial model. Here we simulate Coleman's original model, including
multi-mode spatial effects, although without the gravitational effects
required in a full cosmological theory.

The present paper describes a numerical simulation, whose purpose
is to evaluate the feasibility of an experiment and to predict likely
outcomes. We use the most successful dynamical phase-space representation
for long times, which is the truncated Wigner (tW) approximation to
QFT \citep{Drummond1993,Steel1998}. This uses a $1/N$ expansion
for $N$ bosons, and has had previous success in first-principles
predictions of quantum dynamics in bosonic quantum fields. It gives
correct predictions of tunneling in small bosonic quantum systems
with shallow potentials, by comparison to exact number state and positive-P
representation methods \citep{Kinsler1991}. Cosmological models also
employ relatively flat potentials \citep{Linde1982}, so this is not
unrealistic.

Previous work analyzed quantum bubble nucleation using a $^{41}$K
BEC interferometer \citep{Opanchuk2013,Fialko2015,Fialko2017} at
zero temperature. This proposed a condensate containing atoms of the
same element with two spin components coherently coupled by a microwave
field. The coupled BEC is initialized by a Rabi rotation into a metastable
state, which decays into a stable state through quantum tunneling.
The transition of the BEC from a metastable to a stable state is an
ideal experiment for the investigation of Coleman's original idea.
It simulates the relativistic scalar field as a relative phase between
the spin components, with the speed of light represented by the speed
of sound in the BEC. A related system has been studied using the presence
of a seeded vortex within the condensate to initiate false vacuum
decay \citep{Billam2019_vortex_false_vacuum}.

Here we treat a finite-temperature theory, as will occur in a real
experiment. We utilize a variation of the Bogoliubov method \citep{Bogolyubov194723}
for treating the quantum initial state, which employs a nonlinear
chemical potential to eliminate divergences in the Bogoliubov theory
\citep{Ng2018}. We calculate the effects of thermal noise on vacuum
tunneling using the truncated Wigner approximation, which has given
successful quantum coherence predictions \citep{Drummond1993,Corney:2006_ManyBodyQD,Corney_2008,Egorov_2011,Drummond2017_TWD,opanchuk2019mesoscopic}.
We show that finite temperatures can enhance tunneling, and study
how tunneling rates are modified at different initial laboratory temperatures.
These results include a momentum cutoff to eliminate Floquet instabilities,
and we show that a cutoff is essential by analyzing the effects of
changing the cutoff.

We also treat the dynamics and time-evolution of the observable topological
phase entropy, which has an intuitive, understandable interpretation.
Tunneling events that form the vacuum are disordered, leading to a
peak entropy as time evolves. The entropy nearly reaches the maximum
possible for an appropriate choice of space and phase bins. As a result,
there is a predicted dynamical reduction with time in the topological
entropy, as true vacuum domains are formed. The final vacuum state
is more ordered, since the false vacuum is unoccupied. Domain-wall
formation is minimized at low temperatures, which is important for
cosmological interpretations \citep{zel1974cosmological}, since it
is known that high-temperature domain-wall formation can lead to anomalous
CMB effects. 

\section{Quantum state representation and Hamiltonian\label{sec:Quantum-state-representation}}

\subsection{Field equations}

Bosonic quantum fields with internal degrees of freedom \citep{Guralnik1964,Higgs1964,Englert1964}
are used to describe the Higgs sector in the standard particle model.
Global symmetries of the Hamiltonian are broken while creating the
low-energy ground state or vacuum. The observation of the Higgs particle
makes this an important fundamental concept.

The theory of a metastable or 'false' vacuum was developed by Coleman
\citep{Coleman1,Coleman2}. This used a simpler model, treating the
fundamental quantum dynamics of a scalar quantum field with non-derivative
self-interactions and a Lagrangian density (using a $+---$ metric)
of:
\begin{equation}
\mathcal{L}=\frac{1}{2}\partial_{\mu}\phi\partial^{\mu}\phi-U\left(\phi\right).
\end{equation}

The local field potential was defined to have two spatially homogeneous,
locally stable equilibrium states, $\phi=\phi_{+}$ and $\phi=\phi_{-}$.
The first of these has a higher energy, with $U\left(\phi_{+}\right)>U\left(\phi_{-}\right)$.
This is unstable to quantum corrections, and is expected to decay
to the true vacuum, $\phi_{-}.$ A characteristic predicted to occur
in such decays is that a true vacuum is formed by quantum tunneling
at a particular space-time point, and subsequently grows at the speed
of light.

The dynamics of the evolution of the system is described by Heisenberg
field equations of form: 
\begin{equation}
\partial_{\mu}\partial^{\mu}\hat{\phi}+U'\left(\hat{\phi}\right)=0.\label{eq:relativistic field equation}
\end{equation}
The fact that these are operator equations makes them effectively
insoluble, apart from approximations. Even if all the eigenstates
were known, as in some one-dimensional theories, there are exponentially
many terms \citep{yurovsky2017dissociation} in an expansion of generic
initial states. 

Coleman analyzed a scalar quantum field theory of how such a true
vacuum would arise, given an initial metastable state. This model
also predicted the formation of individual early universe 'bubbles'.
Such theories can be extended to include gravitational effects, and
have been used as a theory of the early universe \citep{Guth_1981}.
In these, the scalar field is renamed the inflaton field, and decay
to a true vacuum causes an inflationary expansion \citep{Linde1982},
creating the 'Big Bang'. More recent cosmological studies often focus
on post-inflationary events \citep{EastherPhysRevLett.124.061301},
which are less sensitive to quantum fluctuations.

The observation of the Higgs particle and evidence for CMB density
fluctuations, confirms the importance of such quantum field theory
models. Yet the original false vacuum energies are thought to be many
orders of magnitude greater than any possible experiment, possibly
$\sim10^{15}\mathrm{GeV}$ \citep{Linde1982}. The original event
is also hidden from direct observation. In addition to such experimental
problems, the quantum field theory itself is exponentially complex,
and cannot be solved exactly. As a result, the theory has mainly been
analyzed using classical or perturbative approximations \citep{mukhanov1992theory}.
The inclusion of general relativistic effects further complicates
the analysis.

It is important to have a better understanding of at least the simplest
quantum field theory models. A feature of the model we use is that
it possesses a spontaneously broken discrete symmetry, which is known
to provide a potential solution to the particle-antiparticle asymmetry
problem \citep{zel1974cosmological}. Qualitative analysis of domain-wall
formation at high initial temperatures has led to objections to this
idea \citep{kibble1980some}, related to CMB spectral observations.
A full quantum dynamical treatment of the location of domain-walls
is needed. Our simulations show that at high temperatures domain walls
are prevalent. However, at low temperatures, domain walls are restricted
to universe boundaries where they appear less likely to cause inhomogeneities
in the CMB spectrum. Observing domain walls in an experimental setting
would help to verify or refute this analysis.

\subsection{Approximations and interpretation }

Since digital quantum computers are orders of magnitude too small,
we propose to solve the equations using an analog quantum computer:
a laboratory BEC quantum simulator. In order to obtain insight into
the expected performance of the simulator, the dynamics of the BEC
system will be solved numerically, but with approximations. Here,
we give an outline of those approximations, explaining where they
will be expected to hold, and also discuss the interpretation of the
simulations.

An interesting consequence of all quantum models for the universe
is that the entire universe is described as a single quantum state:
there is no external observer. Quantum measurement theory in the conventional
Copenhagen model requires an observer to collapse the wave-function.
As a result, there are foundational problems in interpreting the wave-function
itself. This leads to the question of what can one identify in the
simulation that will correspond to a universe? 

The numerical simulations of this paper give predictions of the dynamics
of the wave function for a model of the universe according to a Hamiltonian
treatment. As such the averages taken over the simulations provide
the ensemble predictions for the laboratory experiment if repeated
many times. An interesting question is whether or how a particular
laboratory realization can relate to a particular dynamical trajectory.

We use a mapping of the wave-function to a Wigner field distribution
but with a simpler, approximate time-evolution equation which ensures
a positive Wigner distribution throughout the dynamics. In this truncated
Wigner (tW) model, stochastic equations are written for complex amplitudes
$\alpha_{k}$ that represent modes $k$. These equations are solved
numerically, the quantum noise being modeled stochastically. There
is a direct correspondence between the observable experimental moments
of the field quadratures and the moments of the real and imaginary
parts of $\alpha_{k}$. The measured quantity of interest is the particle
number which, once operator ordering is taken into account, corresponds
to $|\alpha_{k}|^{2}$ up to an error of order $\sim1$. The values
of $|\alpha_{k}|^{2}$ encountered in the simulations are macroscopic,
and hence the difference between operator and simulation moments is
negligible. In this sense, a probabilistic interpretation is possible,
where the individual complex amplitudes trajectories correspond to
an individual realization. Vacuum fluctuations may be considered as
real events, and no additional collapse mechanism is required, as
discussed in greater detail in Section (\ref{sec:Equations-of-motion}).

The dynamics of the tW distribution is approximate. Even though the
local evolution errors when using the tW method are of order $1/N$
where $N$ is the number of particles in each mode, these may grow
during time-evolution to create macroscopic errors at later times
\citep{Drummond1989,Kinsler1991,Sinatra2002,deuar2007correlations}.
Such errors can increase during quantum tunneling. The tW method
cannot describe the formation of macroscopic superposition states
\citep{yurke1986generating,budroni2015quantum,kovachy2015quantum,opanchuk2016quantifying,rosales2018leggett}
, or predict certain macroscopic Bell violations \citep{thenabadu2020testing},
because such states cannot be described by a positive Wigner function.
Quantum squeezing and entanglement can be described however.

As it is a feasibility study, we do not give exact results, because
the quantum simulator experiment is intended to do this. Nevertheless,
it is important to ask how reliable the tW phase-space method is.
The tW distribution has been used to predict both squeezing and entanglement
in systems of large particle number. Comparisons have been made of
tW and exact positive-P methods for the dynamics of quantum squeezing
in solitons \citep{Drummond1993,Bogdan2017-Soliton,Steel1998}, with
excellent agreement between both methods and with experiments \citep{rosenbluh1991squeezed,Corney:2006_ManyBodyQD,Corney_2008}.
There is also good experimental agreement for large-scale quantum
BEC interferometry, where fringe visibility decoherence times have
been accurately predicted. In this regime, quantum entanglement in
the form of Schrodinger's quantum steering has been inferred based
on the simulations, for states of up to 40,000 atoms \citep{opanchuk2019mesoscopic}.

Other studies treated quantum tunneling in driven non-equilibrium
systems, which showed agreement between tW and exact methods for shallow
tunneling potentials \citep{Drummond1989,Kinsler1991}. This agreement
disappeared for deeper potential wells, which is more likely to lead
to macroscopic superposition states requiring negative Wigner distributions.
Another quantum field system with metastable behavior is the quantum
solitonic breather \citep{yurovsky2017dissociation,Bogdan2017-Soliton},
for which tW methods have shown good agreement with conservation laws
\citep{Drummond2017_TWD} and both exact positive-P representation
and integrable methods \citep{ng2019nonlocal,Marchukov2020PhysRevLett.125.050405}
during the early stages of breather relaxation.

While Coleman's original model proposed a deep potential, the models
currently favored by many cosmologists do not. Instead, a shallow
potential is thought to be more realistic in an inflationary universe
\citep{Linde1982}. As a result, it is reasonable to use a relatively
flat quantum field internal potential, with a large particle number.
In this regime, the numerical simulation methods used here appear
reliable. Nevertheless, owing to the long time-scales involved - and
possible error growth \citep{Kupiszewska1990PhysRevA.42.6869,Kinsler1993PhysRevA.48.3310}
- the main goal here is an experiment, regarded as an early universe
quantum simulation.

\subsection{Two-species Hamiltonian}

For a BEC system having two occupied hyperfine levels with mass $m$,
the Hamiltonian of the coupled field system includes an s-wave scattering
potential \citep{Leggett2001RevModPhys.73.307}. It is important in
our model that there is a strong mixing between the two spin species,
without a phase separation. This requires that the inter-species interaction
is minimized, and is assumed to be zero here. There will be losses
due to spin-changing inelastic collisions, but these dissipative effects
are neglected. The size of such effects is not known for the $^{41}$K
Feshbach resonance of interest. This provides a limitation on the
accessible tunneling times, since the atoms must tunnel before they
are absorbed.

A general two-species Hamiltonian includes both intra and inter-species
scattering. The inter-species scattering length is often close to
the intra-species one, since differences in nuclear spin orientation
do not strongly perturb inter-atomic forces. However, this can change
dramatically at a magnetic Feshbach resonance. The required tuning
of the s-wave scattering interactions can therefore be achieved with
the external magnetic field chosen so that cross-species scattering
is suppressed. This is possible in $^{41}\mathrm{}$K, as well as
in other isotopes like $^{7}$Li. 

In these cases, near the Feshbach resonance, one can write the Hamiltonian
as
\begin{equation}
\hat{H}=\stackrel[j=1]{2}{\sum}\hat{H}_{j}+\hat{H}_{c}.\label{eq:Hamiltonian}
\end{equation}
Here, writing $\hat{\Psi}_{j}\equiv\hat{\Psi}_{j}\left(\mathrm{x},t\right)$
for brevity for the $j$-th Bose field, the individual spin-species
Hamiltonians are
\begin{eqnarray}
\hat{H}_{j} & = & \int d\mathrm{x}\left(-\hat{\Psi}_{j}^{\dagger}\frac{\hbar^{2}\nabla^{2}}{2m}\hat{\Psi}_{j}+\frac{g}{2}\hat{\Psi}_{j}^{\dagger2}\hat{\Psi}_{j}^{2}\right),\label{eq:fvac_Ham}
\end{eqnarray}
and the microwave coupling Hamiltonian, $\hat{H}_{c}$, is
\begin{eqnarray}
\hat{H}_{c} & = & -\nu\left(t\right)\int d\mathrm{x}\left(\hat{\Psi}_{2}^{\dagger}\hat{\Psi}_{1}+\hat{\Psi}_{1}^{\dagger}\hat{\Psi}_{2}\right).
\end{eqnarray}

The components $\hat{\Psi}_{j}$ are the coupled field operators corresponding
to different nuclear spin states, and the subscripts $j,k=1,2$ are
the spin indices. These operators have commutation relations $\left[\hat{\Psi}_{j}\left(\mathrm{x},t\right),\hat{\Psi}_{k}^{'}\left(\mathrm{x}',t\right)\right]=0$
and $\left[\hat{\Psi}_{j}\left(\mathrm{x},t\right),\hat{\Psi}_{k}^{'\dagger}\left(\mathrm{x}',t\right)\right]=\delta_{jk}\delta_{M}(\mathrm{x}'-\mathrm{x})$.
Here $\delta_{M}(\mathrm{x}'-\mathrm{x})$ is a restricted delta function
\citep{opanchuk2013functional} that includes a momentum cutoff, restricting
the field to a lattice for numerical simulation. 

The coefficient $g$ is the s-wave scattering interacting strength
between the atoms, which for a three-dimensional system, $g_{3D}$,
is given by:
\begin{equation}
g_{3D}=\frac{4\pi\hbar^{2}a}{m},
\end{equation}
where $a$ is the s-wave scattering length. 

If the atoms are confined by a transverse harmonic trap with frequency
$\omega_{\perp}$, where the transverse trapping energy $\hbar\omega_{\perp}$
is much higher then the thermal energy, the Bose gas reaches a one-dimensional
regime \citep{Deuar2009}. In this regime, the atoms are confined
tightly within an effective s-wave cross-section $A_{s}=2\pi(l_{\perp})^{2}$,
where $l_{\perp}=\sqrt{\hbar/m\omega_{\perp}}$ is the transverse
harmonic oscillator length \citep{olshanii1998atomic,kheruntsyan2005finite,Bouchoule2012}.
The coefficient $g$ for a one-dimensional system is hence expressed
as $g=g_{3D}/A_{s}$, which gives:
\begin{equation}
g=\frac{2\hbar^{2}a}{ml_{\perp}^{2}}=2\hbar a\omega_{\perp}.
\end{equation}

We neglect microwave spontaneous emission effects, as these are very
weak for such microwave transitions.

\subsection{Stephenson-Kapitza pendulum term}

The coupling $\nu$ describes a microwave field that rotates the nuclear
spin by resonantly coupling two hyperfine levels with a frequency
separation of $\Omega_{HF}$ in the external magnetic field. This
is modulated in time in order to induce metastability using Stephenson's
concept of a modulated pendulum \citep{Stephenson1908,stoker1950nonlinear,Kapitza1965_45},
later popularized by Kapitza \citep{Kapitza1965_45}. The depth of
the field potential in the metastable state is determined by the dimensionless
variable $\delta$. 

We work in a rotating frame such that this energy separation is removed
from the Hamiltonian, using a different reference energy for each
spin component. Here $\nu\left(t\right)$, the coupling strength between
the spin components, is modulated \citep{Stephenson1908,stoker1950nonlinear,Kapitza1965_45,Kapitza1965_46,ibrahim2006excitation}
as a sinusoidal time-dependent variable with an additional modulation
frequency $\omega$, so that
\begin{equation}
\nu\left(t\right)=\nu+\delta\hbar\omega\cos\omega t.
\end{equation}

Provided the modulation is at a high frequency, this Hamiltonian is
equivalent to the Coleman model of a relativistic scalar quantum field
with an engineered quartic potential. Here the phonon velocity corresponds
to the speed of light \citep{Opanchuk2013,Fialko2017}. Modulation
amplitudes are relatively large, so that the excitation is nearly
bichromatic \citep{freedhoff1990resonance,zhu1990resonance}, or double-sideband.

The result of including this term is an engineered potential $U\left(\phi_{a}\right)$
for an effective scalar field $\phi_{a}$ which physically is the
phase difference between the two co-existing Bose-Einstein condensates
with phases $\phi_{j}$. We define the atomic phase difference as
$\phi_{a}=\phi_{1}-\phi_{2}-\pi$, so that the false vacuum is at
$\phi_{a}=0$ and the true vacua are at $\phi_{a}=\pm\pi.$ In the
limit of strong particle-particle repulsion, this obeys the relativistic
field equation (\ref{eq:relativistic field equation}), where the
speed of light, $c$, is replaced by the phonon velocity, $c=\sqrt{g\rho_{0}/m}$
in the BEC. 

The potential equation is given by \citep{Fialko2017}:

\begin{equation}
U\left(\phi_{a}\right)=\omega_{0}^{2}\cos\left(\phi_{a}\right)\left[1-\frac{\lambda^{2}}{2}\sin^{2}\left(\phi_{a}\right)\right],
\end{equation}
 where $\omega_{0}=2\sqrt{\nu g\rho_{0}}/\hbar$ and $\lambda=\delta\sqrt{2g\rho_{0}/\nu}.$
This potential can be varied by the experimentalist, and is best understood
in a dimensionless form described later.

It is known that instabilities can form in this model at too low a
modulation frequency. This is due to coupling between effects caused
by the modulation frequency and high-frequency phonon modes \citep{braden2018towards,Braden:2019aa}.
Such effects therefore require the use of high enough modulation frequencies
to move the instability region above any physical cutoff in momentum. 

We will assume that there is a physical mechanism to remove high-momentum
phonon modes. An example of this would be the use of a spatially modulated
potential to introduce a band-gap. A second possibility is the use
of a swept modulation frequency to reduce parametric gain by changing
the unstable momenta. Ultimately, as pointed out by earlier workers,
the inverse scattering length provides an intrinsic cutoff, ultimately
at $1/a$. 

An experimental mechanism to achieve this is via an optically modulated
trap potential. Instabilities were not observed in our previous numerical
simulations, due to the use of a finite lattice that includes a momentum
cutoff. An analysis of modulational instabilities in experiments with
larger numbers of modes and higher phonon momenta is given in the
Appendix, where typical parameters are presented. 

\begin{figure}[h]
\begin{centering}
\includegraphics[width=1\columnwidth]{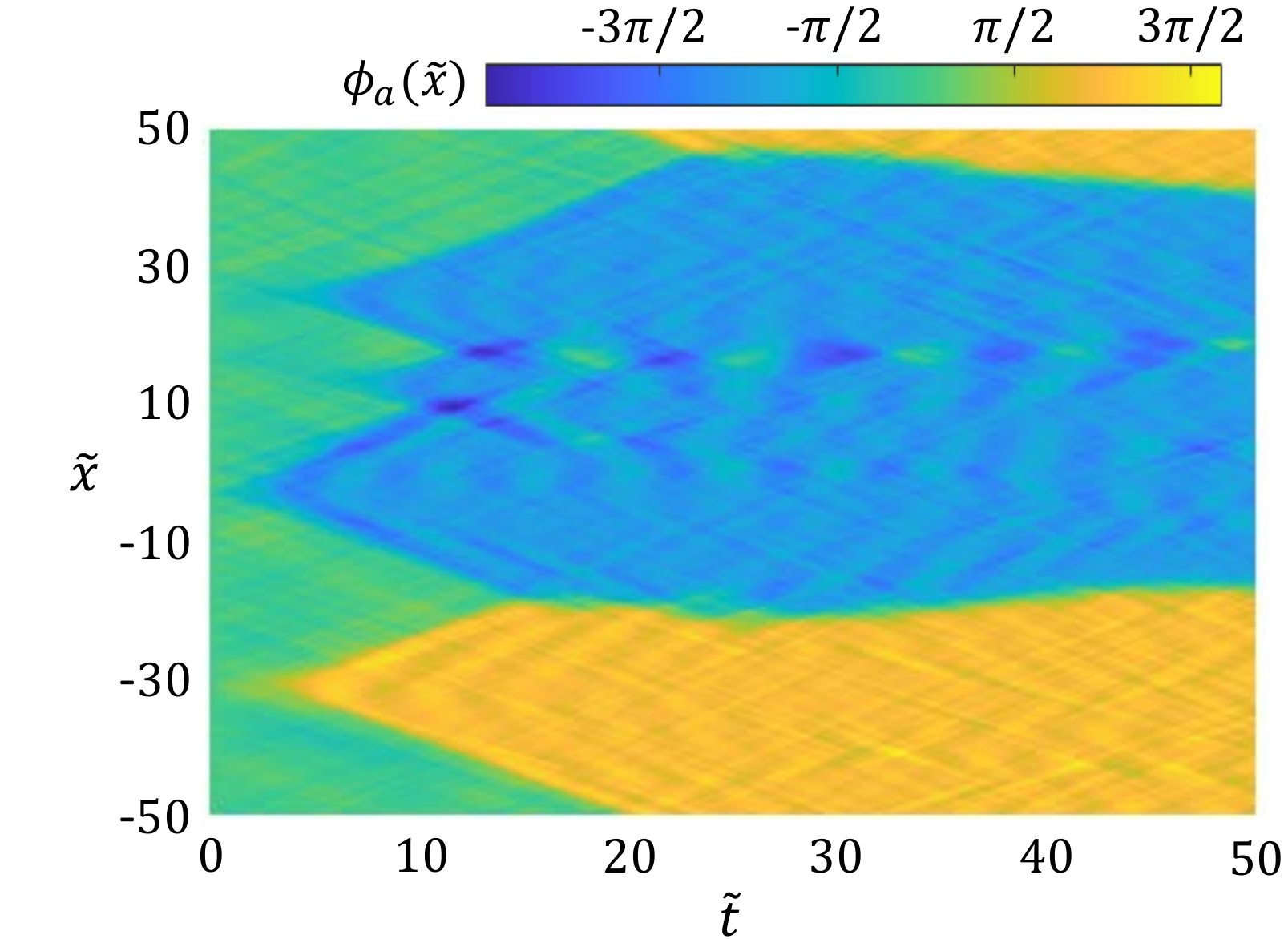}
\par\end{centering}
\caption{Example of a single trajectory decay to a true vacuum, starting in
a $1D$ false vacuum initialized with finite temperature effects.
The false vacuum is seen to decay to two distinct topological phases
for the true vacuum, indicated by yellow and light-blue regions. The
bubbles of true vacuum expand in space until they meet each other.
At long time scales, the universes are separated by domain walls of
false vacuum, indicated by the color green. Dimensionless parameters
are $\tau=10^{-5}$, $\widetilde{\nu}=7\times10^{-3}$, $\lambda=1.2$,
$\widetilde{\omega}=50$ and $\widetilde{\rho}=200$. \label{fig:Phase}}
\end{figure}

We conclude that such instabilities are generally present, but can
be suppressed in the proposed experiment by using high enough modulation
frequencies combined with a momentum cutoff, as shown in Fig (\ref{fig:Phase}).

\section{Initial state at finite temperature\label{sec:Initial-state-at}}

Our initial state includes quantum and thermal fluctuations. We combine
the Bogoliubov method \citep{Bogolyubov194723} with the Wigner representation,
so that both the initial thermal excitations and vacuum noise are
taken into account. In this approximation, the system is assumed to
have a macroscopic condensate mean population $N_{c}$ with density
$\rho_{c}$ for one of the spin indices, say $j=1$. Experimentally,
this is produced using evaporative cooling methods \citep{cornell2002nobel,ketterle1996evaporative,bradley1997bose}.

The ground state field operator $\hat{\Psi}_{1}$ is nearly equal
to the square root of the condensate density, $\sqrt{\rho_{c}}$,
which is assumed constant over the ring. This is valid for typical
ultra-cold BEC experiments below threshold, provided trapping potential
noise is small. There are zero momentum divergences in applying the
Bogoliubov expansion to finite systems, which are removed here using
a nonlinear chemical potential \citep{Ng2018}.

Initially, the microwave coupling $\nu$ is turned off, and the second
spin species operator $\hat{\Psi}_{2}$ is in the vacuum state. The
BEC is prepared in a single species condensate with one spin component
populated at a temperature $T$. While it is possible to achieve temperatures
well below the condensate critical temperature, these are still not
at zero temperature. The condensate has thermal phonon excitations,
which can change the tunneling time. These also induce phase fluctuations
and finite temperatures in the effective scalar field.

This is expected to give modified tunneling compared to a metastable
quantum field without extra noise. Since the exact quantum state prior
to the Big Bang is not known precisely, our goal here is to determine
the effect of thermal noise on a laboratory experiment. Even this
may not capture the full effects of evaporative cooling, which is
a complex dynamical process \citep{drummond1999quantum}.

\subsection{Nonlinear chemical potential }

To model a finite temperature experiment, we assume that the initial
density matrix is in a grand canonical ensemble $\hat{\rho}_{GC}$
at temperature $T$ for $j=1$, and a vacuum state for $j=2$, so
that:
\begin{equation}
\hat{\rho}_{GC}=e^{-\beta\hat{K}}\left|0\right\rangle \left\langle 0\right|_{2}.
\end{equation}
Here $\hat{K}$ is the 'Kamiltonian' , which includes a chemical potential
to give a finite particle number in the thermal state, so that
\begin{equation}
\hat{K}=\hat{H}-\mu\left(\hat{N}\right),
\end{equation}
where $\mu\left(\hat{N}\right)$ is the chemical potential for an
initial population $\hat{N}$ in one of the spin configurations. This
can be any nonlinear function of $\hat{N}$ \citep{Ng2018}, so that,
including terms up to second order
\begin{equation}
\mu\left(\hat{N}\right)=\mu_{1}\hat{N}+\frac{\mu_{2}}{2}:\hat{N}^{2}:.
\end{equation}

The chemical potential has no effect on the dynamics if $\nu=0$,
since $\left[\hat{H},\mu\left(\hat{N}\right)\right]=0$. Such a nonlinear
chemical potential is useful for describing thermal fluctuations in
a BEC. The utility of this approach is that on linearizing the total
Kamiltonian, $\hat{K}$, the zero frequency divergence of the Bogoliubov
expansion \citep{Bogolyubov194723} is eliminated with an appropriate
choice of the quadratic coefficient $\mu_{2}$. 

To allow an expansion around a well-defined coherent state phase,
we suppose that $\hat{\rho}$ is an ensemble average of condensates
$\hat{\rho}\left(\phi_{c}\right)$ with a coherent phase $\phi_{c}$,
so that:
\begin{align}
\hat{\rho}_{GC} & =\frac{1}{2\pi}\int d\phi\hat{\rho}\left(\phi_{c}\right)\,.
\end{align}
This ensemble corresponds to a number-averaged ensemble of states
with Poissonian number fluctuations around $N_{c}$, which is more
realistic than using a fixed particle number. The actual number fluctuations
in experiments are often super-Poissonian, and these additional number
fluctuations can be included if required. 

For number-conserving interferometric measurements of relative phase,
the initial coherent phase $\phi_{c}$ is irrelevant. All phase choices
give identical observables. It is therefore sufficient to include
a single member of the phase ensemble with $\phi_{c}=0$, following
standard methods from laser physics \citep{Glauber1963-states,louisell1973quantum,drummond_hillery_2014}.
There are other techniques for obtaining convergent Bogoliubov expansions
\citep{gardiner1997particle,boccato2019bogoliubov}. These generally
involve operator expansions having non-standard commutators, and do
not readily permit Wigner phase-space expansions.

\subsection{Regularized Bogoliubov theory}

The grand canonical Hamiltonian, $\hat{K}$, is obtained from a Bogoliubov
expansion \citep{Bogolyubov194723,fetter1972nonuniform,lewenstein1996quantum}
of the single-species field operator at time $t=0$, assuming a condensate
phase of $\phi_{1}=0$. This is given by
\begin{equation}
\hat{\Psi}_{1}(x,0)=\psi_{c}+\delta\hat{\Psi}_{1}.
\end{equation}
Here $\psi_{c}=\sqrt{\rho_{c}}$, which is the initial condensate
density, and the field fluctuations are expanded as a sum over phonon
momenta $k$:
\begin{align}
\delta\hat{\Psi}_{1} & =\frac{1}{\sqrt{L}}\underset{k}{\sum}\left[u_{k}\hat{b}_{k}e^{ikx}-v_{k}\hat{b}_{k}^{\dagger}e^{-ikx}\right].\label{eq:Bog_symmetric_breaking}
\end{align}
For a complete unitary transformation, all modes must be included,
including the zero-momentum mode. Next, we introduce the condensate
quadrature operator $\hat{P}$ as
\begin{equation}
\hat{P}\equiv\int d\bm{x}\left(\delta\hat{\Psi}_{1}+\delta\hat{\Psi}_{1}^{\dagger}\right)/\sqrt{2}\,.\label{eq:quadrature}
\end{equation}
The number operator $\hat{N}$ can be written to second order in the
quantum field fluctuations, giving
\begin{equation}
\hat{N}=N_{c}+\hat{P}\sqrt{2\rho_{c}}+\int d\bm{x}\delta\hat{\Psi}^{\dagger}\delta\hat{\Psi}\,.\label{eq:NumberOperator}
\end{equation}

Expanding the grand canonical Hamiltonian in $\delta\hat{\Psi}_{1}$,
the choice of $\mu_{1}=0$ and $\mu_{2}=g/L$ eliminates all first
order terms as well as terms in $\hat{P}^{2}$, which would cause
phase divergences and unphysical phase diffusion in equilibrium. The
resulting convergent Bogoliubov expansion has $u_{0}=1$ and $v_{0}=0$
for the zero-momentum terms with $k=0$, rather than the divergent
expression found in the standard expansion. 

The mode coefficients of the single-species field for $k\neq0$ are
expressed in terms of the excitation energy $\epsilon_{k}$ and the
free-particle energy $E_{k}$, as
\begin{align}
u_{k} & =\frac{\epsilon_{k}+E_{k}}{2\sqrt{\epsilon_{k}E_{k}}},\nonumber \\
v_{k} & =\frac{\epsilon_{k}-E_{k}}{2\sqrt{\epsilon_{k}E_{k}}},\label{eq:v}
\end{align}
where $E_{k}=\hbar^{2}k^{2}/\left(2m\right)$ is the free-particle
energy and
\begin{equation}
{\color{red}{\normalcolor \epsilon_{k}=\sqrt{E_{k}(E_{k}+2g\rho_{c})}}}
\end{equation}
 is the excitation energy. Due to periodic boundary conditions on
a ring trap, the allowed values of momenta are
\begin{equation}
k_{j}=\frac{2\pi j}{L}\,,
\end{equation}
for $j=\left(1-M\right)/2,\ldots\left(M-1\right)/2$ for $M$ momentum
modes, assuming an odd mode number. The resulting effective Kamiltonian
describing thermal excitations of the BEC is given by
\begin{equation}
\hat{K}^{(2)}=\sum\epsilon_{k}\hat{b}_{k}^{\dagger}\hat{b}_{k}\,.
\end{equation}

Here, the phonon operators $\hat{b}_{k}^{\dagger},\hat{b}_{k}$ describe
the creation and annihilation of a quasi-particle in an excited state
$k$. The resulting excitation in each mode with $k\neq0$ is a propagating
phonon. This expansion is particularly useful for the finite temperature
system we are interested in here. Phonon excitations with $k\neq0$
are populated according to the bosonic thermal distribution,
\begin{align}
\langle\hat{n}_{k}\rangle & =\langle\hat{b}_{k}^{\dagger}\hat{b}_{k}\rangle\equiv n_{k}=\frac{1}{\mathrm{exp}(\beta\epsilon_{k})-1},\label{eq:Bog_thermal_distribution}
\end{align}
where $\beta=1/k_{B}T$.

For $k=0$, we assume a vacuum state with $n_{0}=0$, since this operator
cannot be coupled to an energy exchange process, owing to number conservation.
This choice gives Poissonian number fluctuations. As discussed above
it may be necessary in modeling experiments to include larger number
fluctuations due to technical noise occurring in the evaporative cooling
process.

After the initial preparation, a microwave pulse is used to rotate
the Bose gas occupations so that the two spin species have an equal
occupation, which is equivalent to a linear beam-splitter. We denote
$\hat{\Psi}_{j}'(x,0)$ as the initial quantum fields after the BEC
is split into the two states. The two initial spin states after rotation
have equal density $\rho_{0}=\rho_{c}/2$ and a relative phase of
$\pi$.

This corresponds to a rotation matrix acting on the quantum fields
$\hat{\Psi}_{1}$ and $\hat{\Psi}_{2}$:
\begin{align}
\left(\begin{array}{c}
\hat{\Psi}_{1}'\\
\hat{\Psi}_{2}'
\end{array}\right) & =\left(\begin{array}{cc}
\mathrm{cos}\frac{\theta}{2} & -ie^{-i\phi}\mathrm{sin}\frac{\theta}{2}\\
-ie^{i\phi}\mathrm{sin}\frac{\theta}{2} & \mathrm{cos}\frac{\theta}{2}
\end{array}\right)\left(\begin{array}{c}
\hat{\Psi}_{1}\\
\hat{\Psi}_{2}
\end{array}\right),\label{eq:rotation-1}
\end{align}
where $\theta=\pi/2$ and $\phi=-\pi/2$. The system is then assumed
to evolve according to the Hamiltonian (\ref{eq:Hamiltonian}) with
a c.w. microwave coupling field present. This carrier has an appropriate
phase relationship with the microwave pulse field so that the quantum
system is initially in the metastable high energy state.

\subsection{Dimensionless parameters}

The equation of motion and quantum operators can be transformed into
dimensionless form by introducing dimensionless time, distance, and
frequency:
\begin{align}
\widetilde{t} & =t\omega_{0}\nonumber \\
\widetilde{x} & =x/x_{0}=x\omega_{0}/c\nonumber \\
\tilde{\omega} & =\omega/\omega_{0}.
\end{align}
The speed of sound in a weakly interacting BEC is given by $c=\sqrt{g\rho_{0}/m}$,
and the initial temperature of quantum degeneracy is $T_{d}$ \citep{kheruntsyan2003pair}.
We define a characteristic length, temperature and frequency as:
\begin{align}
x_{0} & =\frac{\hbar}{2\sqrt{m\nu}}\nonumber \\
T_{d} & =\frac{\hbar^{2}\rho_{c}^{2}}{2mk_{B}}\nonumber \\
\omega_{0} & =2\frac{\sqrt{\nu g\rho_{0}}}{\hbar}.
\end{align}

The field amplitude in dimensionless coordinates is $\widetilde{\Psi}_{j}=\hat{\Psi}_{j}\sqrt{x_{0}}$,
and the density in dimensionless form is given by $\widetilde{\rho}_{0}=\rho_{0}x_{0}$.
This gives a characteristic energy scale that is the geometric mean
of the mean field energy $g\rho_{0}$ and coupling energy $\nu$,
as used previously. The six dimensionless parameters that define the
physical system are therefore:
\begin{align}
\widetilde{\rho}_{0} & =\rho_{0}x_{0}\nonumber \\
\widetilde{L} & =L/x_{0}\nonumber \\
\tau & =T/T_{d}\nonumber \\
\lambda & =\delta\hbar\omega_{0}/\sqrt{2}\nu\nonumber \\
\widetilde{\omega} & =\omega/\omega_{0}\nonumber \\
\tilde{\nu} & =\nu/g\rho_{0},
\end{align}
where $\widetilde{L}$ is the dimensionless length of the simulation,
$T$ is the temperature of the initial BEC field with density $\rho_{c}=2\rho_{0}$,
and $\lambda$ is the effective depth of the  modulation. The corresponding
dimensionless Hamiltonian, including the chemical potential, is:
\begin{equation}
\tilde{H}_{j}=\int_{-\tilde{L}/2}^{\tilde{L}/2}d\widetilde{x}\left(-\sqrt{\tilde{\nu}}\tilde{\Psi}_{j}^{\dagger}\tilde{\Psi}_{j}+\frac{\tilde{g}}{2}\tilde{\Psi}_{j}^{\dagger2}\tilde{\Psi}_{j}^{2}\right)-\frac{\tilde{g}}{2\tilde{L}}\hat{N}_{1}^{2}.
\end{equation}
Here $\tilde{g}=1/(2\widetilde{\rho}_{0}\sqrt{\widetilde{\nu}})$,
the effective nonlinearity, depends on the other parameters. The chemical
potential has no effect on dynamics, and is only required to remove
singularities in the linearization of the initial state.

\section{Phase-space representations and entropy \label{sec:Equations-of-motion}}

To investigate vacuum nucleation at finite temperature, we take both
quantum and thermal fluctuations into account. This is achieved by
performing stochastic numerical simulations in the Wigner representation
of the full quantum model of the two-component BEC system. These numerical
simulations are not exact, and indeed there is no exact method known.
Such large-scale quantum field calculations are exponentially complex.
This lack of an exact solution is the motivation for a quantum simulation
experiment. However, we can use stochastic methods to investigate
realistic conditions and expected results.

The effect of quantum and thermal fluctuations results in a wide range
of nucleation times. Numerical methods which are not limited to short
simulation time are desirable to simulate the Coleman theory. Here
we choose the truncated Wigner (tW) approach \citep{Wigner1932,Drummond1993,Steel1998}.
It has a sampling error that remains well-controlled over a long simulation
times. This method gives the first quantum corrections in an $M/N$
expansion \citep{Norrie_2006}, were $M$ is the number of modes and
$N$ is the number of atoms. To give reliable results, it is therefore
necessary that $M/N<<1$. This has been confirmed in comparisons with
exact positive-P and complex-P simulations that do not use truncations
\citep{ng2019nonlocal,Drummond1993}.

There have been successful predictions of measured quantum squeezing
in solitons \citep{rosenbluh1991squeezed,Drummond1993-solitons,Drummond1993,Corney:2006_ManyBodyQD,Corney_2008},
propagation effects in BEC lattices \citep{ruostekoski2005dissipative}
and fringe visibility in finite temperature BEC interferometers \citep{opanchuk2019mesoscopic,Steel1998}.
These also indicate that the technique is reliable. Hence the tW method
should be able to simulate these experiments given a large number
of atoms. Truncation errors can build up at long times \citep{Sinatra2002}.
This means that the numerical simulations are expected to be correct
if tunneling is not too slow. The ultimate goal is an experiment. 

Tunneling phenomena are a stringent test of numerical quantum field
simulations. Earlier comparisons of truncated Wigner quantum tunneling
with exact methods have shown that it is correct for relatively shallow
potentials \citep{Drummond1989,Kinsler1991}. This is also the regime
of most interest in many cosmological models. As a result, we expect
that this method will give good indications of the effects of thermal
and quantum noise in the regimes of most interest.

Using this approach, the quantum state is represented by a stochastic
phase space distribution of trajectories following the Gross-Pitaevskii
equation. In a thermal state, the Wigner representation of the initial
state has a complex Gaussian distribution. We perform our simulations
for a one-dimensional system, whose equation of motion are obtained
from (\ref{eq:fvac_Ham}). A typical example is shown in Fig (\ref{fig:Phase}).

\subsection{Wigner representation}

We require that the dynamics of the system is evolved quantum dynamically,
hence quantum fluctuations must be taken into account. To do this,
we transform the phonon operators using the Wigner representation
correspondence \citep{Drummond1993,Steel1998,opanchuk2013functional,Drummond2017_TWD}.
Since the initial state is approximately Gaussian before phase-averaging,
the corresponding initial Wigner representation is also Gaussian \citep{ruostekoski2005dissipative,isella2006quantum}. 

Each initial phonon mode is represented as a complex Gaussian variable,
i.e. $\hat{b}_{k}\sim\beta_{k}$ and $\hat{b}_{k}^{\dagger}\sim\beta_{k}^{*}$.
Thermal fluctuations are included in the modes with $k\neq0$. A detailed
explanation of how this is obtained using the nonlinear chemical potential
method is explained elsewhere \citep{Ng2018}. The Wigner representation
of the initial quantum density matrix $\hat{\rho}\left(0\right)$,
after evaporative cooling, is a complex Gaussian distribution given
by:
\begin{equation}
W\left[\bm{\psi}\right]=W_{1}\left[\psi_{1}\right]W_{2}\left[\psi_{2}\right].
\end{equation}

Here, $W_{1}\left[\psi_{1}\right]$ is a representation of a thermal
state with finite temperature, and $W_{2}\left[\psi_{2}\right]$ is
a vacuum state. These are positive distributions that can be sampled
probabilistically, with samples given after evaporative cooling by:
\begin{align}
\psi_{1} & =\psi_{c}+\frac{1}{\sqrt{L}}\underset{k}{\sum}\left(u_{k}\beta_{k}e^{ikx}-v_{k}\beta_{k}^{*}e^{-ikx}\right)\nonumber \\
\psi_{2} & =\frac{1}{\sqrt{L}}\underset{k}{\sum}\alpha_{k}e^{ikx},\label{eq:wigner-initial-thermal state}
\end{align}
with $\alpha_{k}$ and $\beta_{k}$ defined as independent complex
Gaussian random variables in each vacuum mode and phonon mode respectively.
These have mean values such that $\langle\alpha_{k}^{2}\rangle=\langle\alpha_{k}\rangle=0$,
$\langle\beta_{k}^{2}\rangle=\langle\beta_{k}\rangle=0$. The only
non-vanishing moments are

\begin{eqnarray}
\langle\left|\alpha_{k}\right|^{2}\rangle & = & \frac{1}{2}\nonumber \\
\langle\left|\beta_{k}\right|^{2}\rangle & = & n_{k}+\frac{1}{2}.
\end{eqnarray}

\subsection{Reduction to dimensionless parameters}

In dimensionless form, Eq (\ref{eq:wigner-initial-thermal state})
for the condensate after cooling and before rotation, is given by

\begin{align}
\widetilde{\psi}_{1} & =\widetilde{\psi_{c}}+\frac{1}{\sqrt{\widetilde{L}}}\sum\left(u_{\widetilde{k}}\beta_{\widetilde{k}}e^{i\widetilde{k}\widetilde{x}}-v_{\widetilde{k}}^{*}\beta_{\widetilde{k}}^{*}e^{-i\widetilde{k}\widetilde{x}}\right)\nonumber \\
\widetilde{\psi}_{2} & =\frac{1}{\sqrt{\widetilde{L}}}\sum\alpha_{\widetilde{k}}e^{i\widetilde{k}\widetilde{x}},\label{eq:Wigner-initial-state}
\end{align}
where $|\widetilde{\psi}_{c}|^{2}=\widetilde{\rho}_{c}=2\widetilde{\rho}_{0}$
is the mean field density of the single-species condensate before
rotation. The first term is an inverse Fourier transform of the collective
excitations in Wigner representation. The resulting values for $u,v$
are
\begin{align}
u_{\tilde{k}} & =\frac{\widetilde{\epsilon}_{\widetilde{k}}+\widetilde{E}_{\widetilde{k}}}{2\sqrt{\widetilde{\epsilon}_{\widetilde{k}}\widetilde{E}_{\widetilde{k}}}},\nonumber \\
v_{\tilde{k}} & =\frac{\widetilde{\epsilon}_{\widetilde{k}}-\widetilde{E}_{\widetilde{k}}}{2\sqrt{\widetilde{\epsilon}_{\widetilde{k}}\widetilde{E}_{\widetilde{k}}}},
\end{align}
where $\widetilde{\epsilon}_{\widetilde{k}}=\epsilon_{k}/\hbar\omega_{0}$
is the Bogoliubov excitation energy in dimensionless form, so that,
in dimensionless units:

\begin{align*}
\widetilde{E}_{\widetilde{k}} & =\tilde{\nu}\tilde{k}^{2}\\
\widetilde{\epsilon}_{\widetilde{k}} & =\tilde{\nu}\sqrt{\tilde{k}^{2}\left(\tilde{k}^{2}+\frac{2}{\tilde{\nu}^{2}}\right)}.
\end{align*}

\subsection{Metastable state generation and detection}

Assuming that the thermal phonons effectively behave as a canonical
ensemble of free bosons, the thermal fluctuations for $\widetilde{k}\neq0$
are represented by the complex Wigner amplitude

\begin{align}
\beta_{\widetilde{k}} & =\frac{\eta_{1,\widetilde{k}}}{\sqrt{2\mathrm{tanh}(\widetilde{\epsilon}_{\widetilde{k}}/8\widetilde{\nu}\widetilde{\rho}_{0}^{2}\tau)}},\label{eq:excitation-1}\\
\alpha_{\widetilde{k}} & =\frac{\eta_{2,\widetilde{k}}}{\sqrt{2}},\label{eq:vacuum_noise}
\end{align}
where $\eta_{i,\widetilde{k}}$ is a complex Gaussian noise in dimensionless
space, with $\langle\eta_{i,\widetilde{k}}\eta_{j,\widetilde{k}'}^{*}\rangle=\delta_{ij}\delta_{\widetilde{k}\widetilde{k}'}$
. As a result, 
\begin{align}
\langle|\beta_{\widetilde{k}}|^{2}\rangle & =n_{\widetilde{k}}+\frac{1}{2}\nonumber \\
\langle|\alpha_{\widetilde{k}}|^{2}\rangle & =\frac{1}{2}.
\end{align}

To create the metastable state described in Coleman theory in our
proposed experiment, a radio frequency field with shifted phase of
$\pi/2$ is applied to the single-species BEC. This prepares a superposition
of initial states $|1\rangle$ and $|2\rangle$, where the two-component
condensate corresponds to the initial metastable state, together with
finite temperature thermal fluctuations. 

We denote $\widetilde{\psi}_{1,0}^{'}$ and $\widetilde{\psi}_{2,0}^{'}$
as the initial Wigner fields after the BEC is Rabi-rotated into the
the two hyper-fine levels. These initial states are required to have
equal density $\widetilde{\rho}_{0}$ with a relative phase of $\pi$,
which corresponds to a rotation matrix identical to that used for
the Heisenberg fields, but now applied to the Wigner fields $\widetilde{\psi}_{1}$
and $\widetilde{\psi}_{2}$:

\begin{align}
\left(\begin{array}{c}
\widetilde{\psi}_{1}^{'}\\
\widetilde{\psi}_{2}^{'}
\end{array}\right) & =\left(\begin{array}{cc}
\mathrm{cos}\frac{\theta}{2} & -ie^{-i\phi}\mathrm{sin}\frac{\theta}{2}\\
-ie^{i\phi}\mathrm{sin}\frac{\theta}{2} & \mathrm{cos}\frac{\theta}{2}
\end{array}\right)\left(\begin{array}{c}
\widetilde{\psi}_{1}\\
\widetilde{\psi}_{2}
\end{array}\right),\label{eq:rotation}
\end{align}
where $\theta=\pi/2$ and $\phi=-\pi/2$.

A sample dynamical trajectory in the Wigner phase-space representation
satisfies the equation:

\begin{align}
\frac{\partial\psi_{j}}{\partial t} & =-\frac{i}{\hbar}\left[-\frac{\hbar^{2}\nabla^{2}\psi_{j}}{2m}+g\psi_{j}|\psi_{j}|^{2}-\nu\left(t\right)\psi_{3-j}\right].\label{eq:fvac_equation_motion}
\end{align}

Here we ignore the chemical potential term, which is identical for
both components and has no effect on the relative phase dynamics.
Transforming (\ref{eq:fvac_equation_motion}) into this dimensionless
form, the time-evolution of the Wigner field trajectory is given by
\citep{Fialko2017}:
\begin{align}
\frac{d\widetilde{\psi}_{j}}{d\widetilde{t}}= & -i\left[-\sqrt{\widetilde{\nu}}\widetilde{\nabla}^{2}\widetilde{\psi}_{j}+\tilde{g}\widetilde{\psi}_{j}|\widetilde{\psi}_{j}|^{2}\right]\label{eq:fvac_dimensionless_equation_motion}\\
 & +i\frac{\sqrt{\widetilde{\nu}}}{2}\left(1+\sqrt{2}\lambda\widetilde{\omega}\mathrm{cos}(\widetilde{\omega}\widetilde{t})\right)\widetilde{\psi}_{3-j}.\nonumber 
\end{align}

Increasing the modulation so that $\lambda>1$ gives a local minimum
in the corresponding effective potential. The corresponding dimensionless
effective potential in the phase-difference $\phi_{a}$ is 
\begin{equation}
\tilde{U}\left(\phi_{a}\right)=\frac{4\nu g\rho_{0}}{\hbar^{2}}\cos\left(\phi_{a}\right)\left[1-\frac{\lambda^{2}}{2}\sin^{2}\left(\phi_{a}\right)\right].
\end{equation}

Using the fields $\widetilde{\psi}_{1}^{'}$ and $\widetilde{\psi}_{2}^{'}$
as initial conditions, one can propagate the Wigner fields in real
time, using the equation of motion (\ref{eq:fvac_dimensionless_equation_motion}).
The atomic relative phase then evolves approximately according to
the relativistic field equation (\ref{eq:relativistic field equation}),
so that:

\begin{equation}
\left[\frac{\partial^{2}}{\partial\tilde{t}^{2}}-\widetilde{\nabla}^{2}\right]\phi_{a}+\tilde{U}'\left(\phi_{a}\right)=0.\label{eq:relativistic field equation-1}
\end{equation}
The relative phase of the two spin components is dynamically evolved,
which includes quantum tunneling effects. Since we wish to evaluate
the laboratory experiment, the full atomic equations are evolved,
rather than just the reduced phase equations. We detect vacuum formation
at a finite time by rotating to measure the relative phase from the
resulting hyperfine populations, and the results are compared at different
temperatures. A typical single trajectory is shown in Fig (\ref{fig:Phase}).
This shows tunneling events occurring at isolated space-time points.
As expected, the true vacuum regions grow at the speed of light ($\tilde{c}=1$).
Full details are given in Section (\ref{sec:Numerical-Results}).

\subsection{Topological entropy}

Entropy is an important phenomenon in all physical systems. It is
one of the foundations of thermodynamics, and can be interpreted as
a measure of disorder or randomness of a system. For a quantum system
undergoing unitary evolution, such as the entire universe in this
model, the von Neumann entropy is invariant. While von Neumann entropy
can increase when the contributions for different entangled parts
of the universe are summed, the overall von Neumann entropy for the
universe is static. Alternative definitions of entropy have emerged
that are based on measurement properties \citep{Jaynes1964-entropy,Wehrl1978-Entropy,Swendsen,Balibrea2015-Entropy,Goldstein-conference}.
This leads to the question of which entropy measure can be used to
quantify the disorder of an early universe simulator, and how it can
be calculated and measured.

Here we investigate the disorder of the simulated universe using an
observational macroscopic entropy that can be calculated and measured.
This is based on a well-known quantum entropy measure, the Wehrl entropy
\citep{Wehrl1979-entropy}, 
\begin{equation}
S_{Q}=-\int Q(\bm{{\alpha}})\ln\left(Q(\bm{{\alpha}})\right)d^{2M}\bm{{\alpha}},
\end{equation}
where $Q(\bm{\alpha})=\left\langle \bm{\alpha}\right|\hat{\rho}\left|\bm{\alpha}\right\rangle /\pi^{M}$
is the Husimi function \citep{Husimi1940-Qfunction}. The Q function
is a positive, probabilistic representation, defined for all quantum
states. It can be used to link the simulations and an interpretation
of the quantum universe, based on the Q function \citep{drummond2020retrocausal}.
In this interpretation, the universe simply corresponds to a particular
sample of a Q-function probability.

It is nontrivial to simulate the Q-function dynamics, since it does
not satisfy a Fokker-Planck equation. Consequently, rather than solving
for the Q-function directly - which would be equivalent to quantum
field dynamics - we have utilize a closely related method, the truncated
Wigner (tW) approximation \citep{Wigner1932,Drummond1993,Steel1998}.
This has much simpler dynamical equations. 

Measurements that correspond to a Q-function trajectory are anti-normally
ordered, so that averages over the symmetrically-ordered tW simulations
do not directly correspond to those of a Q-function trajectory. These
two distributions corresponding to two different representations,
one approximate, the other accurate. In any experiment on a large
BEC, the difference between the truncated Wigner and the Q distribution
is microscopic, and has a negligible effect on macroscopic observables
like the average phase difference. Since the ordering introduces
differences of only a microscopic order, either can have the interpretation
of macroscopic reality \citep{reid2017interpreting}. The retrocausal
nature of the individual trajectories and their relationship to quantum
measurement theory is treated in detail elsewhere \citep{drummond2020retrocausal}.
In this interpretation,  no additional collapse mechanism is required.

If we consider a mesoscopic observable, one can approximate the Q-function
distribution by the Wigner function used here, since the two are related
by a microscopic convolution of order $\hbar$ \citep{Glauber1969-convolution}:
\begin{equation}
Q(\bm{{\alpha}})=\frac{1}{\pi^{M}}\int W(\bm{{\alpha'}})e^{-2\left|\bm{\alpha'}-\bm{\alpha}\right|^{2}}d^{2M}\bm{\alpha'}.
\end{equation}
However, even the Wehrl entropy measure, though simpler than the von
Neumann entropy, is not readily measurable in a multi-mode system,
as it requires an exponentially complex set of measurements. To resolve
this problem, the idea of an observational entropy has been recently
put forward, which uses a finite set of measurements to define entropy
\citep{Dominik2020-ObservationalEntropy}.

We use an observational version of the Wehrl entropy, which is a combination
of the Wehrl and observational entropies. Each amplitude $\bm{\alpha}$,
describing a possible universe in the Wigner or Q-representation,
is reduced to a phase, measured and binned into a set $S_{i}$ which
classifies phases into $p$ distinct ranges in each of $\ell$ contiguous
regions. This is a topological measurement, since the phase can only
be established through a nonlocal phase-unwrapping algorithm \citep{Itoh1982-PhaseUnwrapping},
which allows one to distinguish topological phases of $-\pi$ and
$\pi$ through continuity in space and time.

Given $n_{i}$ as the number of measured universes in the $i^{th}$
bin, from a total of $n=p^{\ell}$, we define a probability $P_{i}=n_{i}/n$,
and a corresponding observational Wehrl entropy as
\begin{equation}
S_{T}=-\sum_{i}P_{i}\ln P_{i}\le\ell\ln p.
\end{equation}
An early universe simulation must have a finite number of trajectory
samples. Hence, we require an appropriate binning strategy to formulate
sample probabilities, in which the total number of bins should be
less than the number of samples, to give reliable estimates. The simplest
strategy would be to use a binary binning with the relative phase
at each point in space in either a false vacuum or in a true vacuum,
thus ignoring topological effects. 

However, as shown in Fig (\ref{fig:Phase}), if we start with a false
vacuum at $\phi=0$, we find two topologically distinct true vacuum
states with relative phases of $-\pi$ and $\pi$. These are distinguishable
using nonlocal phase unwrapping methods in space or time. As a result,
we require $3$ phase bins in each spatial region, of $-\pi\pm\pi/2$,
$0\pm\pi/2$ and $\pi\pm\pi/2$  to capture the vacuum states in
our early universe model. Since phase is measured in finite volume,
we define it by averaging over a range of neighboring spatial lattice
points. Therefore we identify in each space-time interval three phase
bins, two for the true vacua, and one for the false vacuum. Compared
to entropy as a microscopic quantity, this entropic measure is uniquely
sensitive to macroscopic topological disorder. In fact, it is sensitive
to disorder on the scale of the entire universe, or model universes
in the case of the proposed laboratory quantum simulations using coupled
Bose condensates. This has quite different properties to the microscopic
von Neumann entropy.

Our simulations show a cooperative effect, where each vacuum bubble
eventually becomes dominated by one or other of the topological phases.
This provides a model for multiple universes with fundamentally different
properties. An intriguing property of this type of scalar field symmetry
breaking is that it is purely topological. There are no local measures
that distinguish the different topological phases, although they are
distinguishable using phase unwrapping. 

It is speculated that discrete symmetry breaking could provide a mechanism
for matter-antimatter asymmetry \citep{zel1974cosmological,kibble1980some}.
The basis for this is that scalar field behavior involves very high
energies, with matter and antimatter being formed at much lower energies.
As a result, small asymmetries could have a large influence at the
lower energies of matter formation. The validity of this model hinges
on the question of whether domain walls form, as these can alter the
observed CMB spectrum. Our simulations indicate that domain wall formation
is suppressed at low temperatures and confined to universe boundaries.
Experimental evidence of domain wall formation and symmetry breaking
can be found by measuring the topological entropy.

\section{Numerical Results\label{sec:Numerical-Results}}

This section will summarize the results of numerical studies of the
effect of finite initial temperatures in the proposed BEC experiments.
Our simulations use a discrete lattice, corresponding to a physical
momentum cutoff at $M=256$ modes. This is necessary to prevent modulational
instabilities. The effect of removing the momentum cutoff by using
a smaller lattice spacing with more modes is reported in the Appendix.
It is experimentally challenging to measure relative phase unambiguously
in a BEC, as this requires a simultaneous measurement of two complementary
quadratures. Most of the numerical results presented here use the
more accessible measure of relative number distribution, $p_{z}\propto\cos\left(\phi_{a}\right)$,
while we also present results for the relative phase $\phi_{a}$ in
the section on topological phase entropy.

\subsection{Experimental parameters}

To have a realistic numerical study, we have chosen possible parameters
that correspond to a one-dimensional system of $^{41}$K atoms in
a ring trap near a Feshbach resonance. There are many choices of atomic
species possible, including $^{7}$Li, so this is only one scenario
among many.

For the existence of quasi-particles in a finite temperature condensate,
the circumference $L$ of the trap should be shorter than the temperature-dependent
phase coherence length $l_{\phi}\approx\frac{\hbar^{2}\rho_{o}}{mk_{B}T}$
\citep{kheruntsyan2003pair}. The restriction on the trap circumference
$L\ll l_{\phi}$ limits the temperature $T$ of the condensate in
the ring trap, i.e. $T\ll T_{c}=\frac{\hbar^{2}\rho_{c}}{mk_{B}L}$
\citep{Fialko2017}. For condensates in a ring trap with a three-dimensional
density $\rho_{c,3D}$, assuming the condensate atoms are transversely
confined in the effective s-wave scattering cross-section $A_{s}$,
the corresponding one-dimensional density is estimated to be $\rho_{c}=N_{c}/L\approx A_{s}\rho_{c,3D}$.
Following the suggested parameters in \citep{Fialko2015,Fialko2017},
the parameters of the proposed experiments are listed in Table \ref{tab:parameters}.

\begin{table}[h]
\begin{tabular}{|c|c|}
\hline 
Experimental parameters & \tabularnewline
\hline 
\hline 
Trap circumference $L$ & $254\mathrm{\mu\mathrm{m}}$\tabularnewline
\hline 
Number of atoms $N_{c}$ & $4\times10^{4}$\tabularnewline
\hline 
Condensate density $\rho_{c}$ & $\approx1.58\times10^{8}\mathrm{m}^{-1}$\tabularnewline
\hline 
Degeneracy temperature $T_{d}=\frac{\hbar^{2}\rho_{c}^{2}}{2mk_{B}}$ & $\approx147\mathrm{\mu\mathrm{K}}$\tabularnewline
\hline 
Coherence temperature $T_{c}=\frac{\hbar^{2}\rho_{c}}{mk_{B}L}$ & $\approx7.34\mathrm{n\mathrm{K}}$\tabularnewline
\hline 
BEC temperature $T$ & $\approx1.47\sim14.7\mathrm{n\mathrm{K}}$\tabularnewline
\hline 
Transverse frequency $\omega_{\perp}$ & $2\pi\times1910\mathrm{Hz}$\tabularnewline
\hline 
Oscillator frequency $\omega$ & $2\pi\times9.56\mathrm{kHz}$\tabularnewline
\hline 
Oscillator amplitude $\nu/\hbar$ & $2\pi\times9.56\mathrm{Hz}$\tabularnewline
\hline 
s-wave scattering strength $g$ & $8.05\times10^{-39}\mathrm{J\cdot m}$\tabularnewline
\hline 
Effective s-wave scattering cross-section $A_{s}$ & $8.10\times10^{-13}\mathrm{m}^{-2}$\tabularnewline
\hline 
3-dimensional condensate density $\rho_{c,3D}$ & $1.94\times10^{20}\mathrm{m^{-3}}$\tabularnewline
\hline 
Speed of sound $c$ & $3.05\times10^{-3}\mathrm{ms^{-1}}$\tabularnewline
\hline 
Observation time $t_{f}$ & $49.9\mathrm{ms}$\tabularnewline
\hline 
Characteristic length $x_{0}$ & $2.54\mathrm{\mu m}$\tabularnewline
\hline 
Characteristic frequency $\omega_{0}$ & $2\pi\times191.26\mathrm{Hz}$\tabularnewline
\hline 
\end{tabular}

\caption{\label{tab:parameters}Dimensional parameters in the proposed experiments. }
\end{table}

The partial differential equations (\ref{eq:fvac_dimensionless_equation_motion})
are solved using an interaction picture fourth-order Runge-Kutta (RK4)
method with the extensible open source MATLAB package xSPDE \citep{Kiesewetter2016xspde}.
From the experimental parameters listed in \ref{tab:parameters},
the corresponding typical dimensionless parameters used in the numerical
simulations are listed in Table \ref{tab:dimensionless_parameters}.

\begin{table}[h]
\begin{tabular}{|c|c|}
\hline 
Typical parameters & \tabularnewline
\hline 
\hline 
Dimensionless circumference $\widetilde{L}$ & $100$\tabularnewline
\hline 
Dimensionless observation time $\widetilde{t}_{f}$ & $60$\tabularnewline
\hline 
Number of modes $M$ & $256$\tabularnewline
\hline 
Dimensionless lattice spacing $\Delta\widetilde{x}$ & $0.3906$\tabularnewline
\hline 
Dimensionless time-step $\Delta\widetilde{t}$ & $7.5\times10^{-4}$\tabularnewline
\hline 
Reduced temperature $\tau$ & $10^{-5}\sim10^{-4}$\tabularnewline
\hline 
Dimensionless atom density $\widetilde{\rho}_{0}$ & $200$\tabularnewline
\hline 
Dimensionless coupling $\widetilde{\nu}$ & $0.004\sim0.01$\tabularnewline
\hline 
Dimensionless modulation $\lambda$ & $1.2\sim1.4$\tabularnewline
\hline 
Dimensionless frequency $\widetilde{\omega}$ & $50\sim200$\tabularnewline
\hline 
\end{tabular}

\caption{\label{tab:dimensionless_parameters}Typical dimensionless parameters
in the numerical simulations.}
\end{table}

\subsection{Observational criteria\label{subsec:Observational-criteria}}

In order to convert the relative phase of the two species into number
density distribution, a $\pi/2$ radio frequency pulse can experimentally
be applied to the coupled fields. Vacuum nucleation can be observed
from the relative number density distribution,

\begin{align}
p_{z}(\widetilde{x}) & =\frac{\rho_{2}(\widetilde{x})-\rho_{1}(\widetilde{x})}{\rho_{2}(\widetilde{x})+\rho_{1}(\widetilde{x})},\label{eq:fvac_pz}
\end{align}
where $\rho_{1}(\widetilde{x})$ and $\rho_{2}(\widetilde{x})$ are
the number density of the two species respectively, after applying
the second Rabi rotation.

\begin{figure}[h]
\includegraphics[scale=0.48]{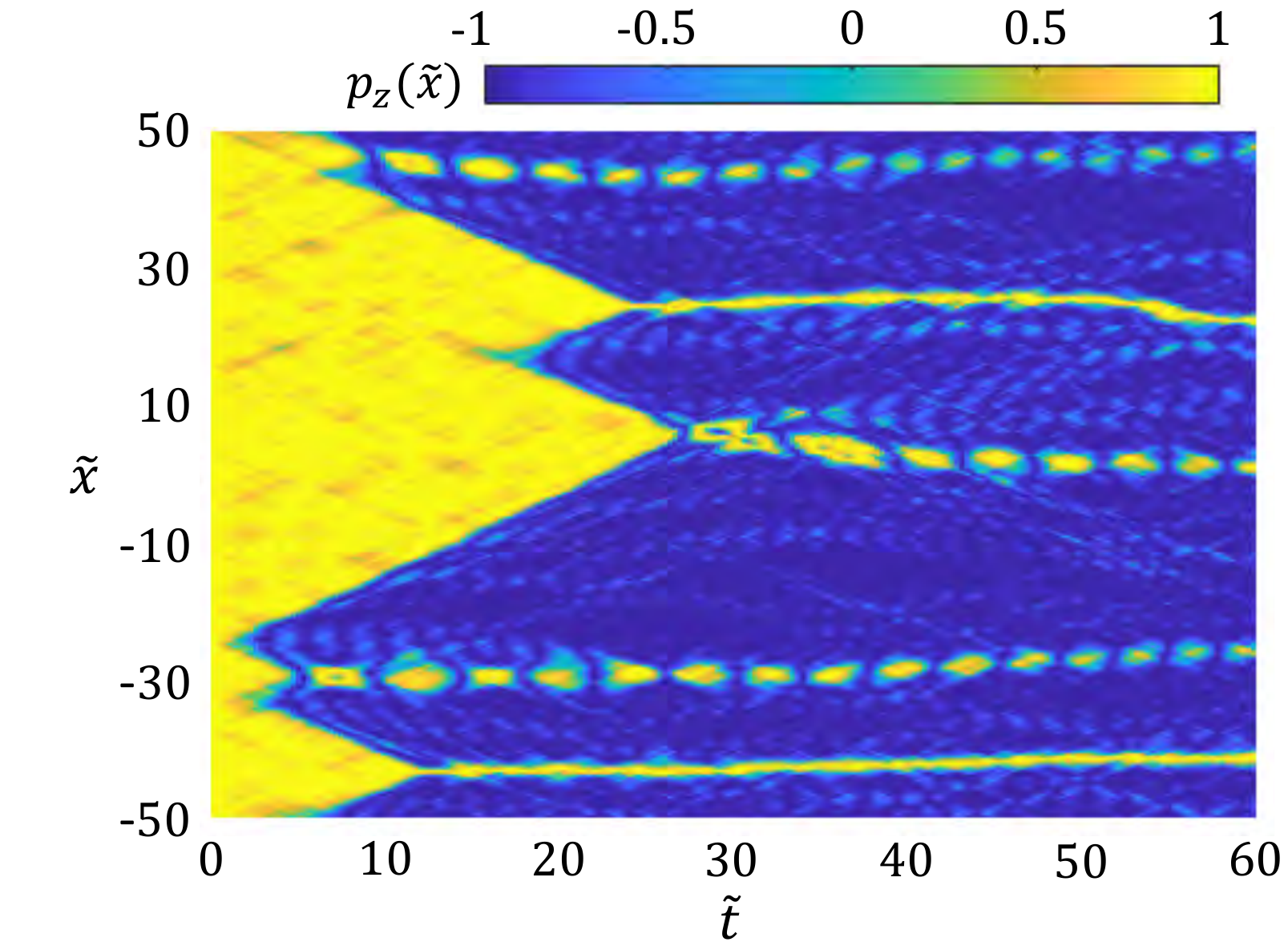}

\caption{\label{fig:tau1e_5}Single-trajectory 1D false vacuum simulation for
the time evolution of $p_{z}$ at $\tau=1\times10^{-5}$. The dimensionless
length $\tilde{L}=100$ corresponds to a trap circumference $L=254\mathrm{\mu\mathrm{m}}$.
Simulation parameters are: $\lambda=1.2$, $\widetilde{\nu}=7\times10^{-3}$,
$\widetilde{\omega}=50$, $\widetilde{\rho}=200$, number of modes
$M=256$. The false vacuum ($p_{z}=1$) indicated by the yellow color
decays, forming bubbles in the true vacua ($p_{z}=-1$) indicated
by the blue regions.}
\end{figure}

Figure \ref{fig:tau1e_5} shows a single trajectory example of one-dimensional
false vacuum dynamics. The simulation starts with thermal states of
a two-component condensate at a low reduced temperature of $\tau=1\times10^{-5}$.
The coupled field system is in the metastable state initially, with
$p_{z}\sim1$ at time $\widetilde{t}=0$, indicated by the yellow
contour. The system starts to decay into a stable true vacuum state
with $p_{z}\sim-1$, indicated by the blue contour at times $\widetilde{t}\gtrsim2$.
In this example, five bubbles are formed of true vacua. These bubbles
expand until they either meet at continuous domain walls of false
vacuum (at $\widetilde{x}\approx-43$ and $\widetilde{x}\approx25$),
which correspond to topologically distinct phases, or else form localized
oscillons (at $\widetilde{x}\approx-30$, $6$ and $46$).

From the simulations of our model at finite temperature, the thermal
energy introduces extra thermal fluctuation into the system. The influence
of the thermal fluctuation can result in an increase of tunneling
events within the coupled BEC fields.

\begin{figure}[h]
\includegraphics[scale=0.48]{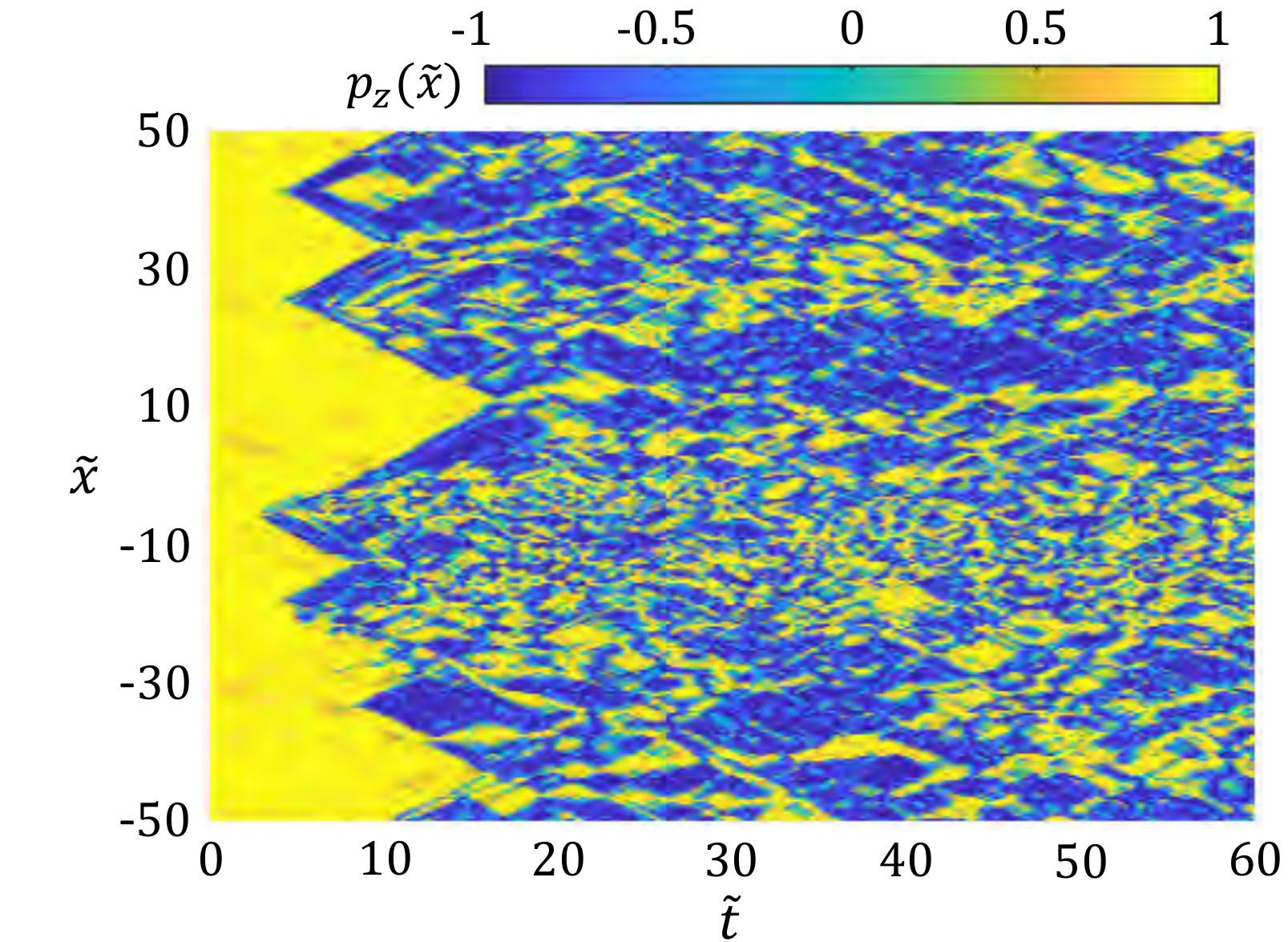}

\caption{\label{fig:tau3e_4}Single-trajectory 1D false vacuum simulation for
the time evolution of $p_{z}$ at $\tau=3\times10^{-4}$, all other
parameters are as in Figure \ref{fig:tau1e_5}.}
\end{figure}

In addition, as shown in the single trajectory example in Figure \ref{fig:tau3e_4},
increasing the reduced temperature $\tau$ of the system enhances
the interactions between the false vacua and the true vacua on long
time scales. In the example of the low temperature dynamics shown
in Figure \ref{fig:tau1e_5}, domain-wall and oscillon formation is
minimized and the bubbles are well defined. This clear structure of
the true vacuum is disturbed when the BEC is strongly thermalized
as shown in Figure \ref{fig:tau3e_4}.

As can be seen in Figure \ref{fig:tau3e_4}, no stable domain walls
or periodic oscillons are formed as the temperature increases, and
no true vacuum bubbles survive on long time scales. As quantum tunneling
is enhanced at higher temperatures, more bubbles are formed in the
true vacua, but most of them are short-lived. Tunneling is accelerated
at high temperatures, which leads to strong fluctuations between the
true vacua and the false vacua, and many relatively unstable domain
walls.

\begin{figure}[h]
\includegraphics[scale=0.48]{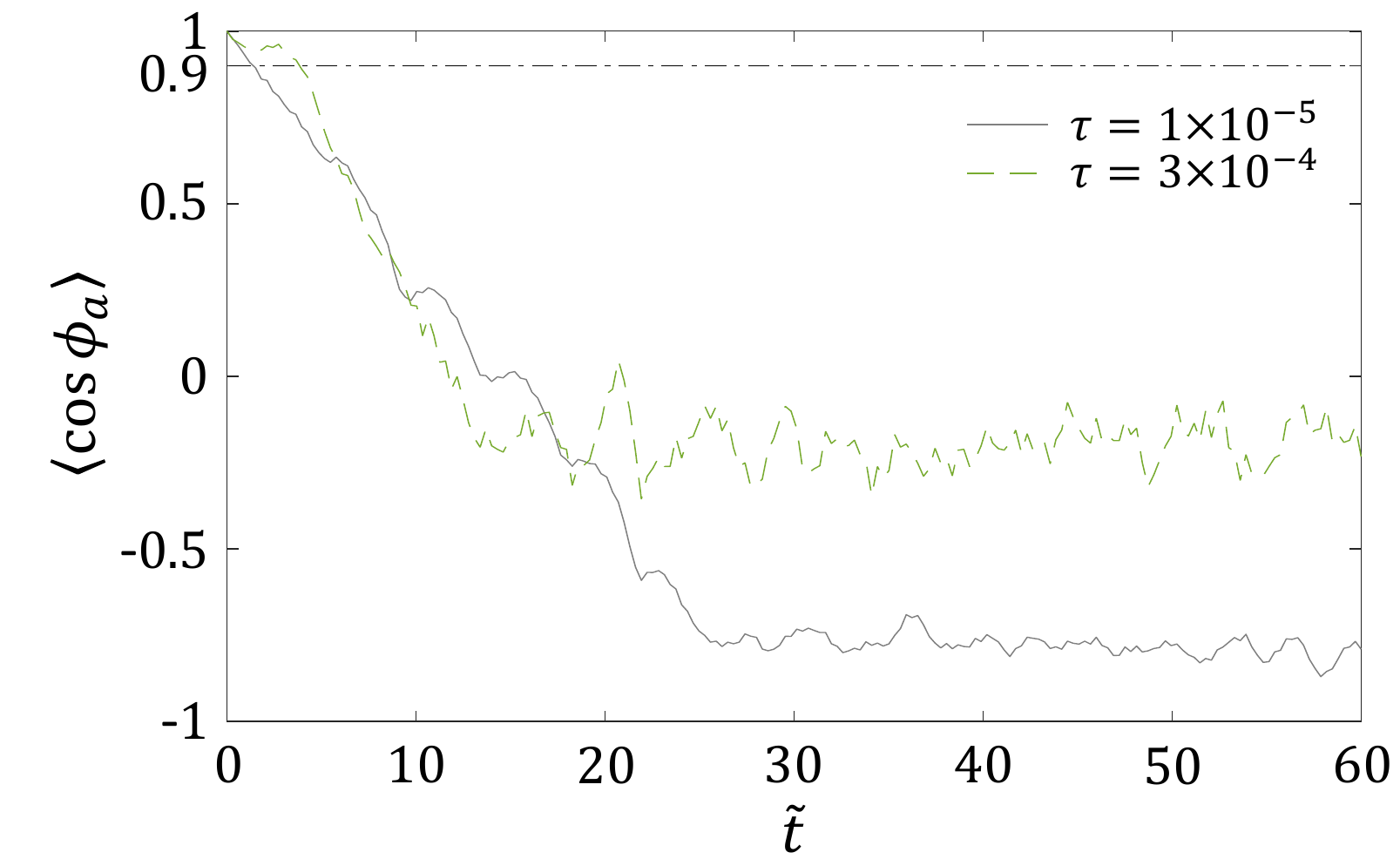}

\caption{\label{fig:tau1e_5_phase}Time evolution of the corresponding average
relative phase $\langle\mathrm{cos}\phi_{a}\rangle$ shown in Figure
\ref{fig:tau1e_5} (solid line) and Figure \ref{fig:tau3e_4} (dash
line), the horizontal dash-dot line $\langle\mathrm{cos}\phi_{a}\rangle=0.9$
indicates the threshold of the appearance of a true vacuum bubble.}
\end{figure}

To quantify tunneling events, we can examine the average relative
phase of the coupled fields along the axial coordinate, where

\begin{align}
\langle\mathrm{cos}\phi_{a}\rangle & =\frac{1}{\widetilde{L}}\intop^{\widetilde{L}}\mathrm{cos}\phi_{a}(\widetilde{x})d\widetilde{x}.\label{eq:average_phase}
\end{align}
As illustrated in Figure \ref{fig:tau1e_5_phase}, the relative average
phase $\langle\mathrm{cos}\phi_{a}\rangle=1$ corresponds to an initial
false vacuum. As the tunneling starts and the false vacuum decays
to the true vacua, the average relative phase is expected to gradually
decrease from $1$ to $-1$ in a complete transition. At very low
temperatures, the presence of the true vacuum bubbles is noticeable
with $\langle\mathrm{cos}\phi_{a}\rangle<-0.5$. However, at higher
temperatures, the presence of the true vacuum bubbles is less noticeable
due to the influence of the thermal fluctuations, and $\langle\mathrm{cos}\phi_{a}\rangle$
only goes to just below $0$. We define a threshold value $\langle\mathrm{cos}\phi_{a}\rangle=0.9$
as the initiation of the false vacuum tunneling event.

Recalling that the tW method provides quantum estimation from a set
of stochastic trajectories, hence to investigate the thermal effect
on the tunneling rate at finite temperature conditions, we repeat
the single trajectory simulation and determine the probability of
bubble creation ($\langle\mathrm{cos}\phi_{a}\rangle\leq0.9$) over
time $P(\widetilde{t})$.

We then obtain the survival probability of the false vacuum $\mathcal{F}$
given that $\mathcal{F}=1-\int_{0}^{\widetilde{t}}P(\widetilde{t}')d\widetilde{t}'$.
The tunneling rate $\Gamma$ for long time scales is calculated from
a linear fit in log scale given that $\mathcal{F}=\mathrm{exp}(-\Gamma\widetilde{t})$
\citep{Takagi}. We performed simulations over a range of coupling
strengths $\widetilde{\nu}$ and reduced temperatures $\tau$. Results
are presented in Figure \ref{fig:FVac lambda1_2}, \ref{fig:FVac lambda1_3}
and \ref{fig:FVac lambda1_4}. Previous work showed that larger couplings
$\widetilde{\nu}$ extend the tunneling time \citep{Fialko2015,Fialko2017},
which confirmed an expected slowing-down of quantum tunneling with
dissipation \citep{Leggett1981_dissipation_tunneling}. To illustrate
thermal effects, we compare the Bogoliubov thermal state results with
the coherent state results which thermal noise is neglected. Although
the coherent initial state is not a true Bogoliubov ground state,
it has very similar behavior to the low temperature ground state.
In the Wigner representation, the coherent initial state is represented
by adding the quantum noise in each vacuum mode (\ref{eq:Wigner-initial-state})
to the classical false vacuum state. In dimensionless from, the coherent
Wigner fields after the BEC is Rabi-rotated are $\widetilde{\psi}_{i=1,2}=\text{\ensuremath{\frac{\widetilde{\psi_{c}}}{\sqrt{2}}}}+\frac{1}{\sqrt{\widetilde{L}}}\sum\alpha_{\widetilde{k}}e^{i\widetilde{k}\widetilde{x}}$,
where $\alpha_{\widetilde{k}}$ is an independent Gaussian random
variable as already described in (\ref{eq:vacuum_noise}).

\subsection{Thermally induced changes in decay rates\label{subsec:Thermally-induced-changes}}

\begin{figure}[h]
\begin{centering}
\includegraphics[scale=0.3]{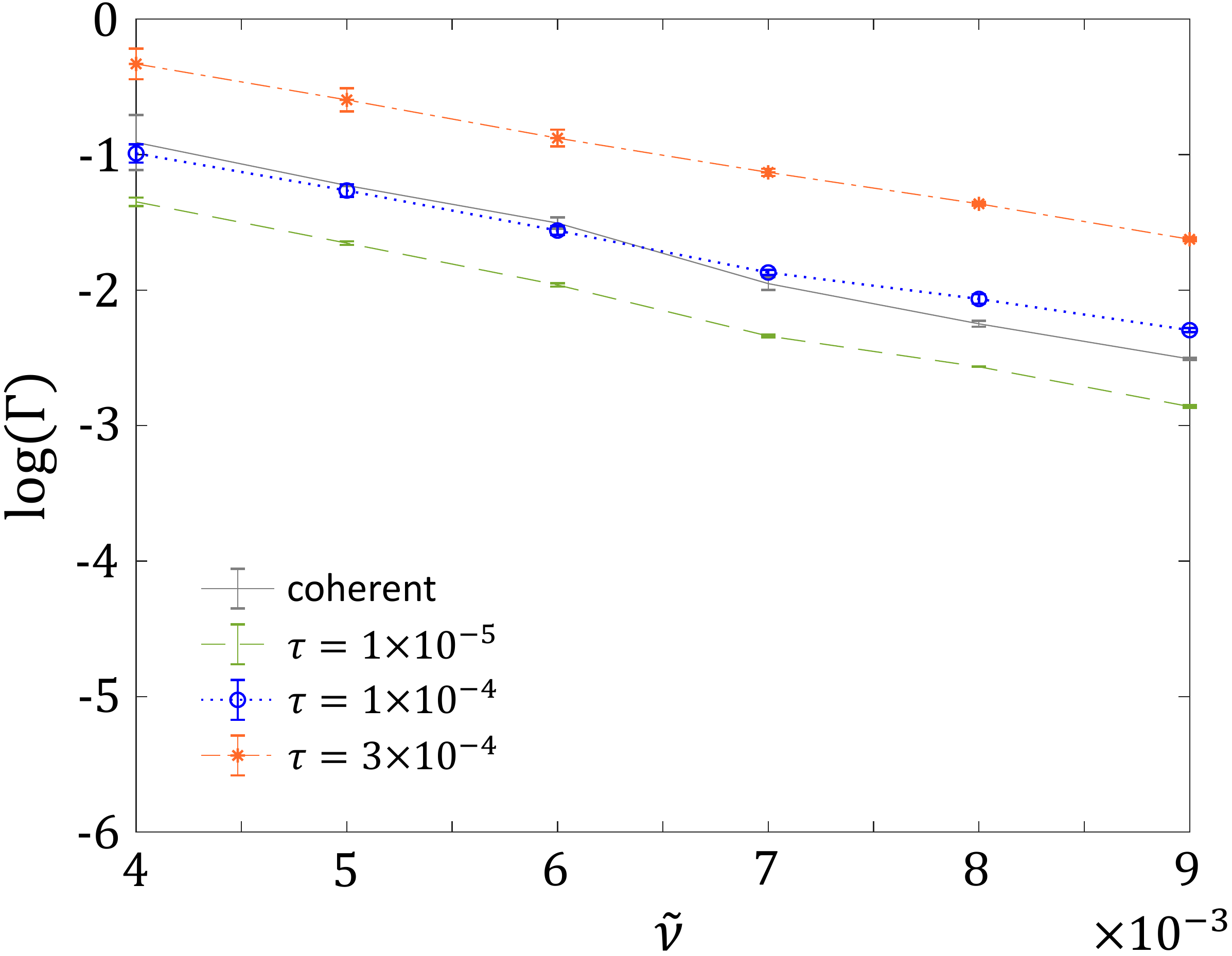}
\par\end{centering}
\caption{\label{fig:FVac lambda1_2}Dependence of the tunneling rate $\Gamma$
on the coupling $\widetilde{\nu}$ for different values of reduced
temperature $\tau$ for $\lambda=1.2$. The error-bars show the estimated
error of the linear least squares fitting in log scale.}
\end{figure}

\begin{figure}[h]
\begin{centering}
\includegraphics[scale=0.3]{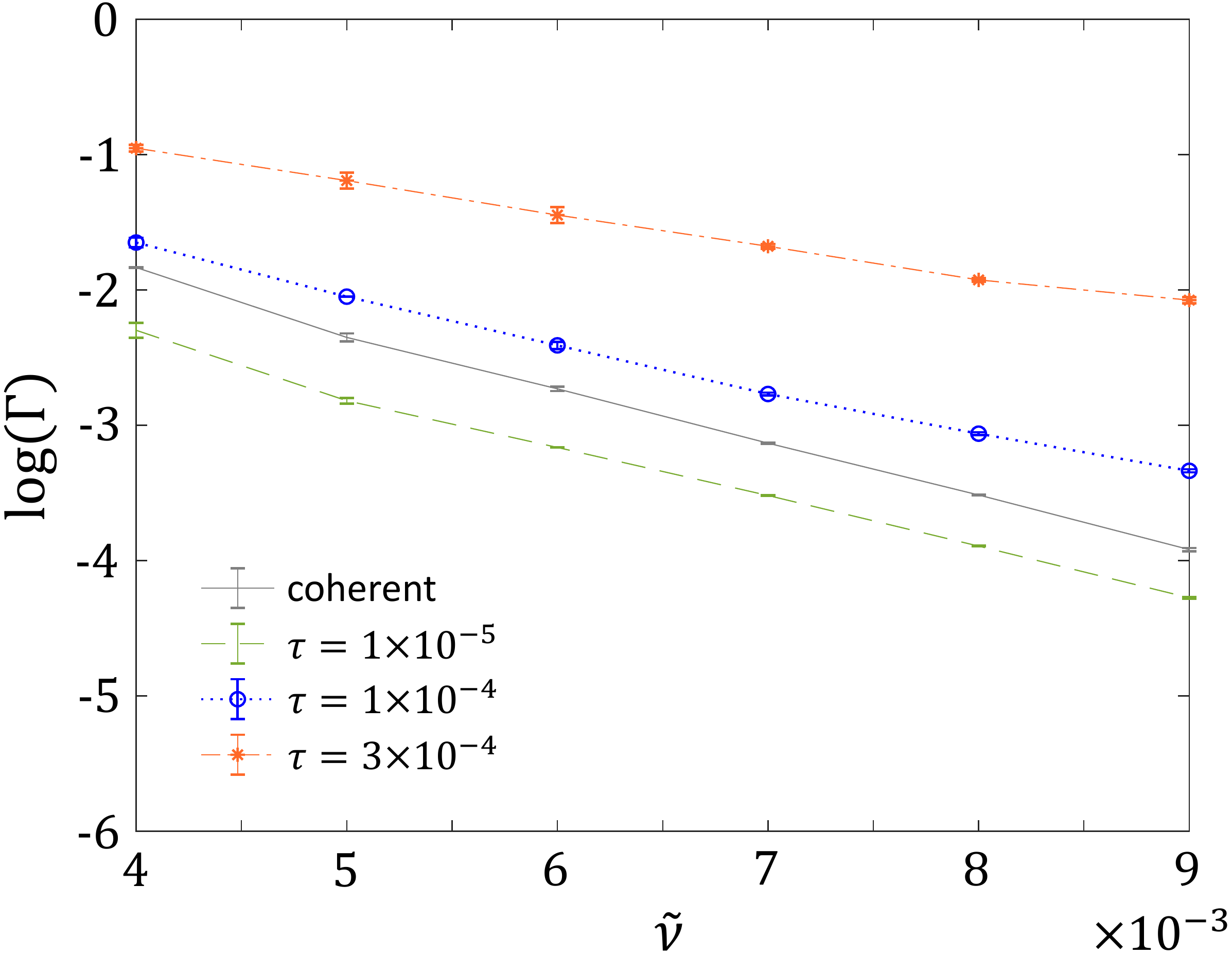}
\par\end{centering}
\caption{\label{fig:FVac lambda1_3}Tunneling rate $\Gamma$ for different
values of reduced temperature $\tau$ for $\lambda=1.3$.}
\end{figure}

\begin{figure}[h]
\begin{centering}
\includegraphics[scale=0.3]{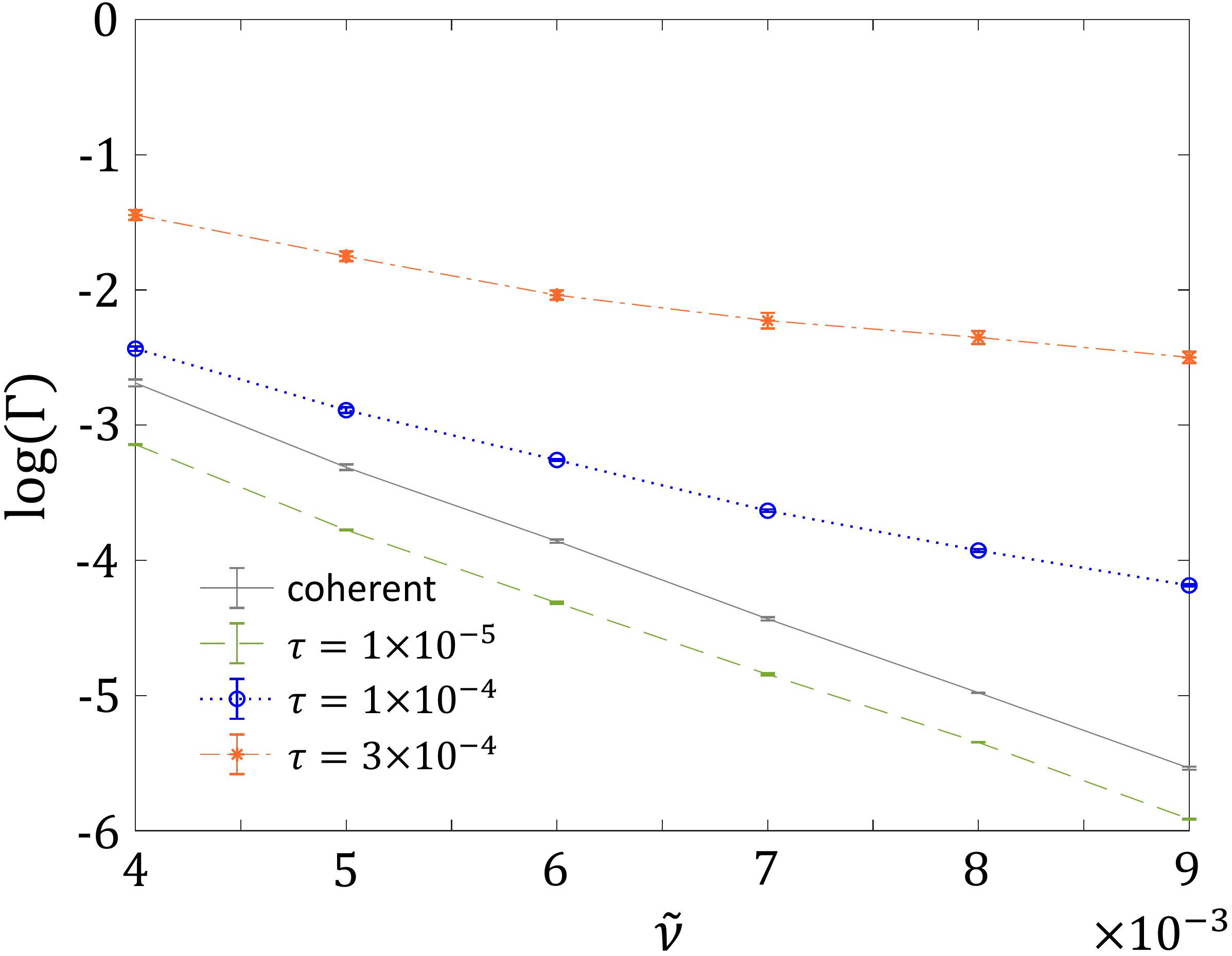}
\par\end{centering}
\caption{\label{fig:FVac lambda1_4}Tunneling rate $\Gamma$ for different
values of reduced temperature $\tau$ for $\lambda=1.4$.}
\end{figure}

How does the finite initial temperature affect the decay of the false
vacuum? From all three figures (Figure \ref{fig:FVac lambda1_2} to
\ref{fig:FVac lambda1_4}), the tunneling rates $\Gamma$ determined
from both the coherent state and the thermal states at all tested
temperatures $\tau$ show a power law dependence on $\widetilde{\nu}$.
The gradients of $\log\Gamma$ for each effective modulation depth
$\lambda$ are similar. If we compare the change of the tunneling
rate $\Gamma$ at a fixed modulation depth $\lambda$, from each of
Figure \ref{fig:FVac lambda1_2}, \ref{fig:FVac lambda1_3} and \ref{fig:FVac lambda1_4},
one can see that the rate of tunneling is increased as the temperature
$\tau$ increases. 

For the case of very low temperatures, $\tau=1\times10^{-5}$, as
the modulation depth $\lambda$ increases the tunneling rate is reduced
more significantly than at higher temperatures. On the other hand,
in the case of the highest reduced temperature studied ($\tau=3\times10^{-4}$),
this reduction of the tunneling rate due to an increase in $\lambda$
is less significant than at lower temperatures. The tunneling of the
false vacuum is less restricted by $\lambda$ at higher temperatures.

This result suggests that at high temperatures, the effect of thermal
fluctuations dominates the quantum decay of the false vacuum. For
a fixed coupling strength $\widetilde{\nu}$, an increase of temperature
increases the probability of penetrating the modulation depth barrier
$\lambda$, and hence increases the formation rate of the true vacuum.
The correlated ground state with $\tau=10^{-5}$ has a slightly lower
tunneling rate than a coherent state, as it is stabilized by the Bogoliubov
correlations. The qualitative effects of bubble nucleation and growth
are similar in both cases.

\subsection{Topological entropy results}

\begin{figure}[h]
\begin{centering}
\includegraphics[width=1\columnwidth]{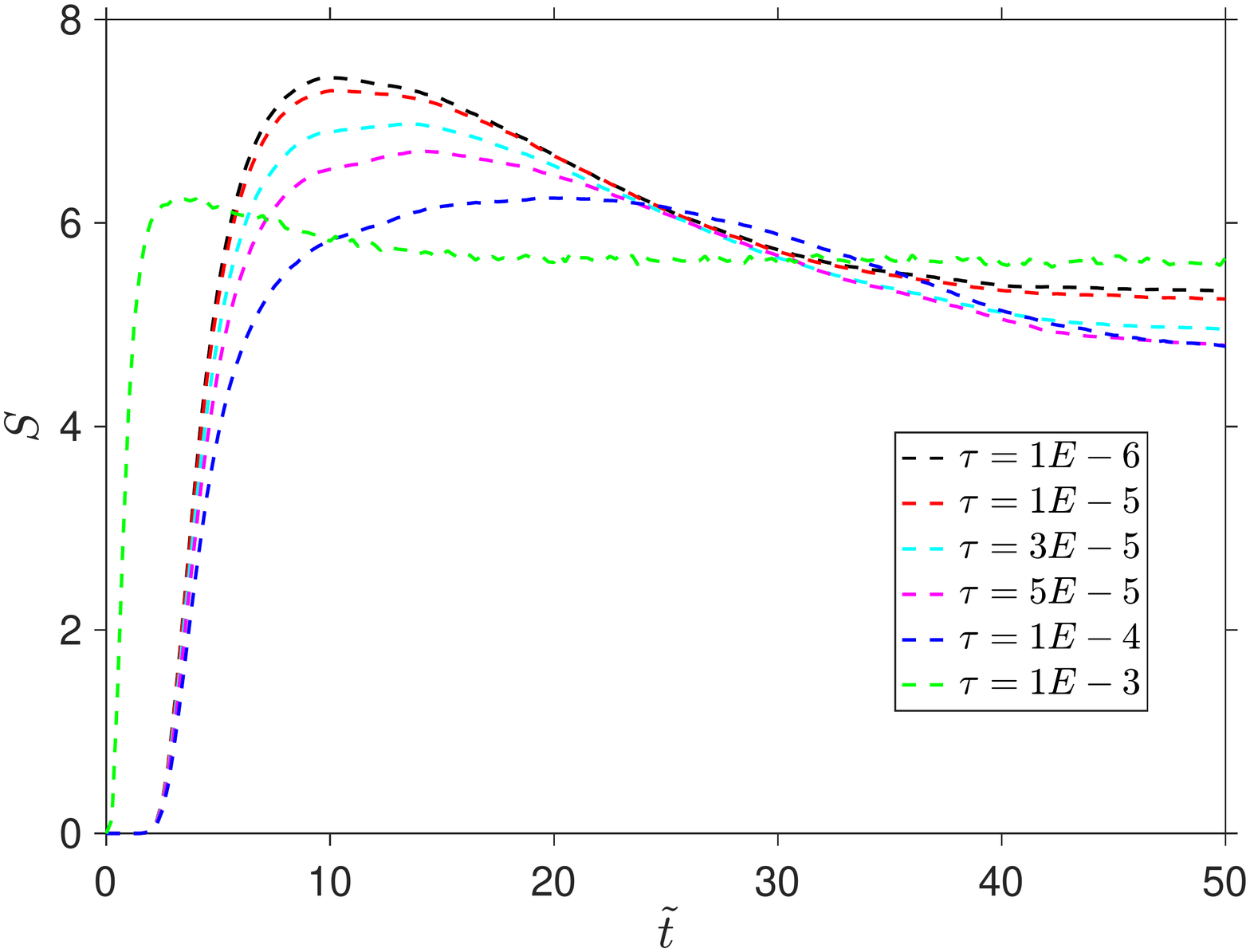}
\par\end{centering}
\begin{centering}
\par\end{centering}
\caption{Evolution of the topological phase entropy function $S$ in early
universe simulations using $10^{4}$ trajectories with varying order
of magnitude, of reduced temperature $\tau$. Here the selected dimensionless
parameters are $\widetilde{\nu}=7\times10^{-3}$, $\lambda=1.2$,
$\widetilde{\omega}=50$ and $\widetilde{\rho}=200$ with simulation
parameters $\widetilde{L}=100$, space-phase bins $\ell=8$, $p=3$,
with $N_{grid}=300$ and $8\times10^{4}$ time steps. \label{fig:S_tau}}
\end{figure}

The topological entropy allows us to quantify the phase disorder caused
by the formation of true vacua and domain walls. Results for the topological
entropy with $10^{4}$ distinct realizations are shown in Fig (\ref{fig:S_tau}).
Initially the coarse grained observational entropy is nearly zero,
since all phase-space coordinates are in the metastable false vacuum.
The observational entropy initially increases with time evolution,
in contrast to the von Neumann quantum entropy which is constant with
time. The entropy reaches a maximum as tunneling occurs, giving a
nearly maximally disordered state with entropy $S<8\ln3=8.79$, using
$\ell=8,$ and $p=3.$ The entropy then reduces as the true vacua
grow, reducing phase disorder.

The entropy function is shown to reach a maximum as a result of tunneling
and after $\widetilde{t}\gtrsim40$, it stabilizes to near $S\sim8\ln2$,
as the false vacuum is eliminated. Between $\tau=10^{-6}$ and $\tau=10^{-4}$
shown by black and red curves, there is a decrease in the maximal
entropy $S_{max}$ and lifetime of the peak. From $\tau=10^{-5}$
to $10^{-4}$, as $S_{max}$ decrease the lifetime of the peak increase.
Beyond $\tau=10^{-4},$ the system thermalizes, as shown by the green
curve.

At very low temperatures, the final state has almost no false vacuum.
As a result, $S\le5.55$ as seen in the simulations of Figure \ref{fig:S_tau}.
At higher temperatures the false vacuum never disappears completely,
due to unstable domain wall formation, thus increasing the final state
disorder. This is shown by the red and black scattered lines in Figure
\ref{fig:S_tau} and leads to the prediction of stable steady-state
entropy values $S\sim5.3$. The remaining disorder arises from the
randomness due to the two topological phases of the true vacua and
the remaining false vacuum domain wall or oscillating oscillons. The
amount of disorder possible is limited by the spatial extent chosen
for phase bins, and the number of samples used.

Figure \ref{fig:Phase} reveals true vacua states after a time $\widetilde{t}\gtrsim3$,
showing two different colors that depict the two different topological
phases for the true vacuum. The true vacuum bubbles expand in space
until they meet after a time $\widetilde{t}\gtrsim15$ at $\widetilde{x}\approx-19$
and after a time $\widetilde{t}\gtrsim24$ at $\widetilde{x}\approx46$.
The neighboring true vacuum regions with distinct topological phases
are separated by a domain wall of false vacuum. On long time scales,
Figure \ref{fig:Phase} clearly depicts the formation of an asymmetric
structure of the topological phase in the true vacuum.

\begin{figure}[h]
\begin{centering}
\includegraphics[width=1\columnwidth]{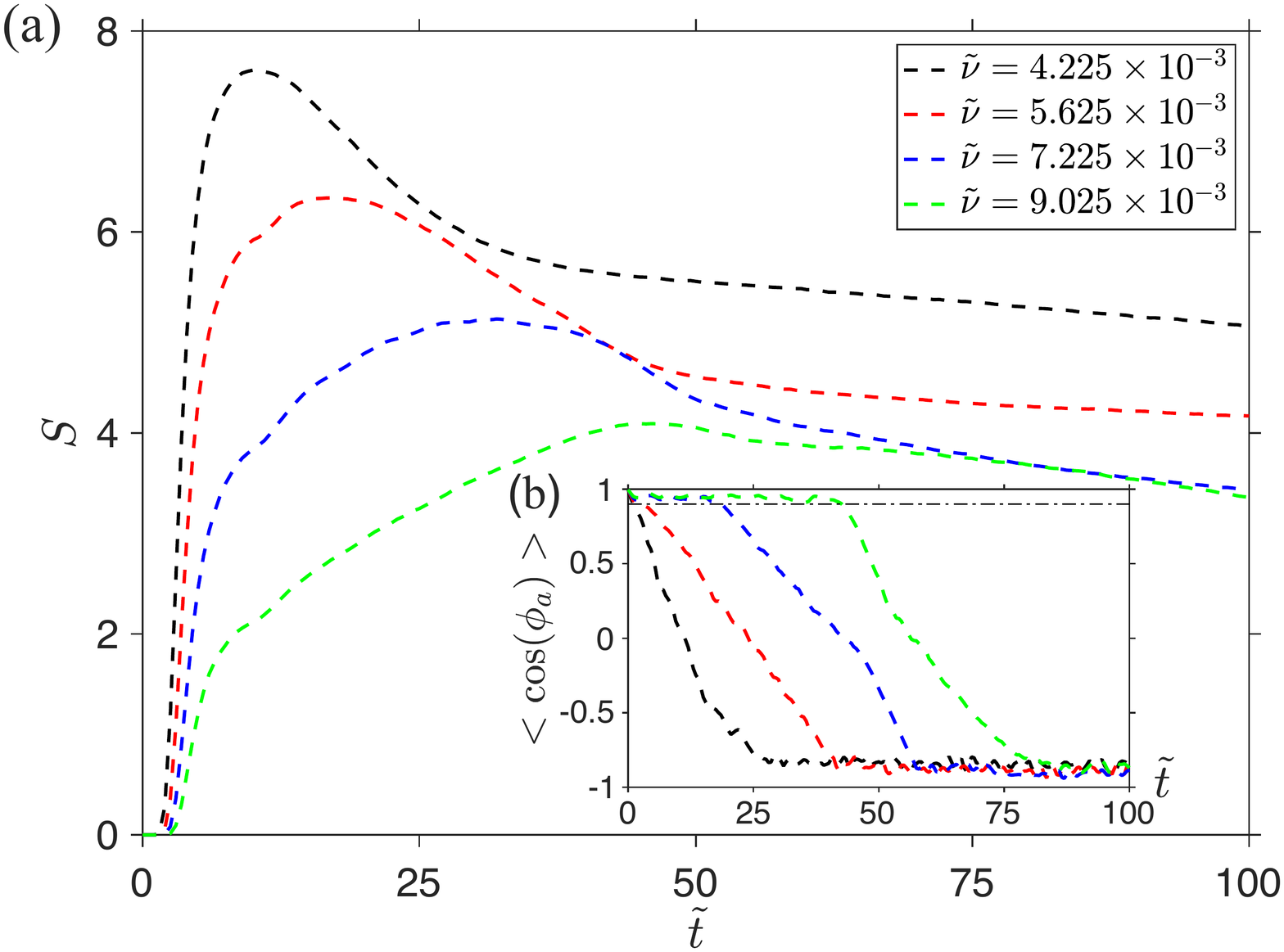}
\par\end{centering}
\caption{(a) Dependence of the entropy function $S(\widetilde{t})$ on coupling
$\widetilde{\nu}$ calculated using $10^{4}$ trajectories with $\lambda=1.3$,
$\widetilde{\omega}=50$, $\widetilde{\rho}=200$ and $\tau=10^{-5}$.
(b) Average relative phase $\left\langle \cos(\phi_{a})\right\rangle =\frac{1}{\widetilde{L}}\intop_{0}^{\widetilde{L}}\cos\phi_{a}(\widetilde{x})d\widetilde{x}$
for single trajectory simulations for different values of $\widetilde{\nu}$.
 The horizontal dash-dot line $\left\langle \cos(\phi_{a})\right\rangle =0.9$
indicates the threshold of the appearance of a true vacuum bubble.
The tunneling time is longer for larger couplings $\widetilde{\nu}$.
\label{fig:S_nu} \textcolor{blue}{}}
\end{figure}

In Figure \ref{fig:S_nu} we display the evolution of the entropy
function $S(\widetilde{t})$ for different values of $\widetilde{\nu}$
using $10^{4}$ tW trajectories, and a comparison plot for the evolution
of average relative phase $\langle\mathrm{cos}(\phi_{a})\rangle$
of the corresponding single trajectory examples. For larger couplings
$\widetilde{\nu}$, the maximum entropy $S_{max}$ is reduced and
the lifetime of the peak is extended. As the coupling $\widetilde{\nu}$
is increased, the tunneling initiation is generally delayed and the
tunneling time is extended. 

The single trajectory results presented in Figure \ref{fig:S_nu}
(b) are consistent with the expectation of a slowing-down of tunneling.
We find similar behavior for the modulation depth $\lambda$ in the
regime of metastability (for $\lambda>1$). In the case of large $\lambda$,
bubble nucleation is delayed, with more results given in the Appendix.

\begin{figure}[h]
\begin{centering}
\includegraphics[width=1\columnwidth]{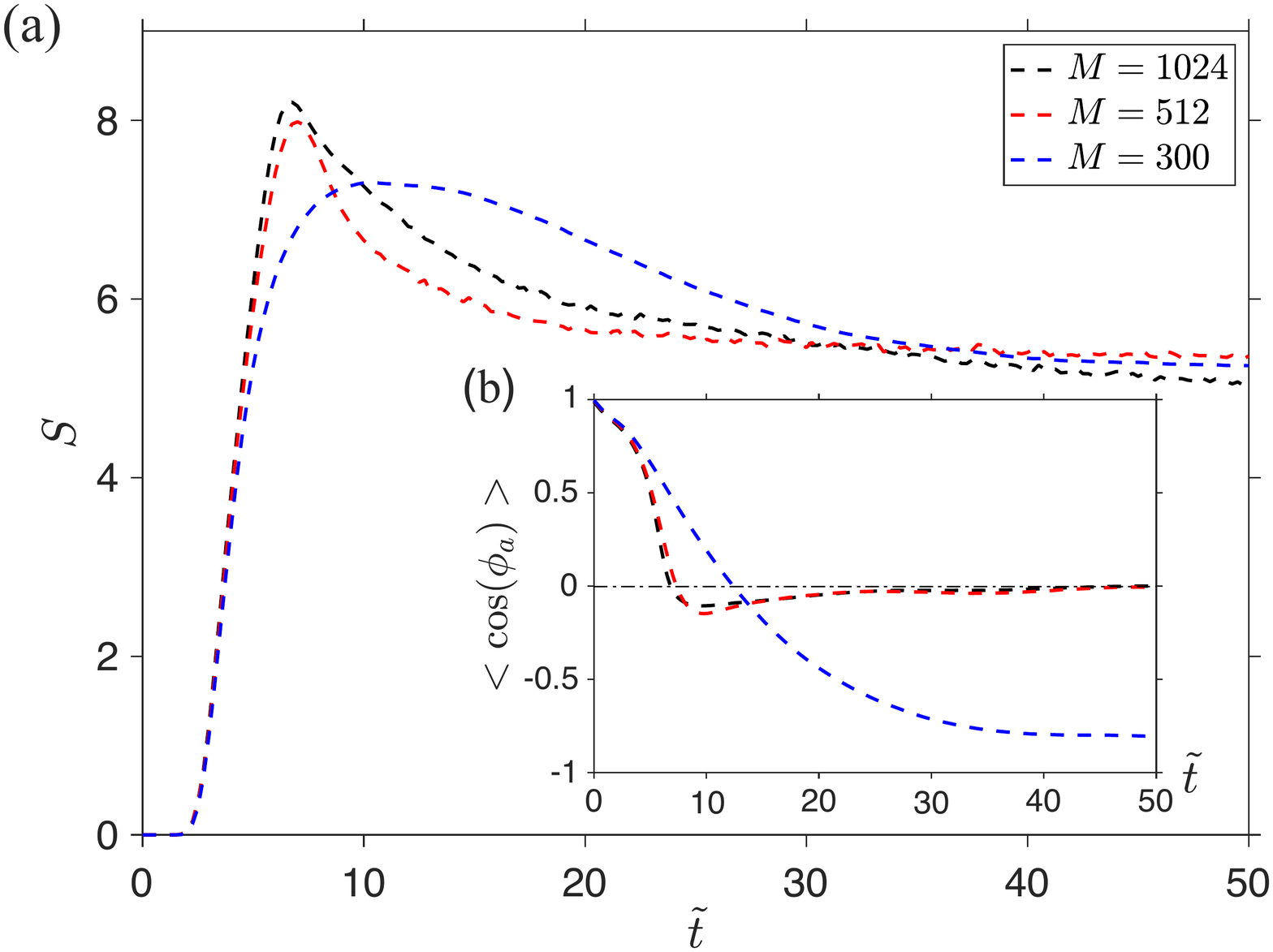}
\par\end{centering}
\caption{(a) Time evolution of the entropy $S(\widetilde{t})$ using $10^{4}$
trajectories for number of modes $M=300$, $512$ and $1024$. The
dimensionless parameters chosen here are $\widetilde{\omega}=50$,
$\tau=10^{-5}$, $\widetilde{\nu}=7\times10^{-3}$, $\lambda=1.2$
and $\widetilde{\rho}=200$. (b) Time evolution of the average relative
phase for different numbers of modes $M$. The horizontal dash-dotted
line indicates a uniform random phase. \label{fig:S_M}  }
\end{figure}

The time evolution of the entropy function $S(\widetilde{t})$ for
different numbers of spatial modes $M$ using $10^{4}$ trajectories
is shown in Figure \ref{fig:S_M}. The results are compared with the
corresponding average relative phase $\langle\mathrm{cos}(\phi_{a})\rangle$
shown in Figure \ref{fig:S_M} (b). The evolution of $S(\widetilde{t})$
and $\langle\mathrm{cos}(\phi_{a})\rangle$ for higher spatial mode
numbers are specified by black and red scattered lines in Figure \ref{fig:S_M}.
Systems with higher mode numbers have a more chaotic phase fluctuation,
whereas the system with lower mode numbers specified by the blue dashed
line show a smooth tunneling to the true vacuum. 

The behavior is reflected in the entropy plots in Figure \ref{fig:S_M},
where the narrow peaks of the entropy function for black and red scattered
lines indicate a more chaotic phase ordering. On the other hand, the
flattened peak of the entropy function for blue scattered line shows
a relatively stable phase. In our simulations, the rapid transition
to chaotic fluctuations at higher spatial modes can be minimized by
increasing the characteristic frequency to $\widetilde{\omega}=200$
to achieve a stable tunneling to true vacuum. This behavior is caused
by Floquet mode instabilities, and is an artifact of the microwave
modulation frequency, as explained in greater detail in the Appendix.

\section{Conclusion}

Prior to the present paper, the proposed experiment on the decay of
a false vacuum had only been studied previously in the zero-temperature
limit \citep{Fialko2017,Opanchuk2013,braden2018towards}. As a result
the effects of finite temperature thermal noise on vacuum tunneling
and nucleation using BECs with two spin components were not understood.
This mixture of two BEC spin components can be used as the relativistic
analogous quantum field, where the relative phase of the two species
corresponds to the metastable state (phase $\pi$) and stable state
(phase zero) respectively. The components can be coupled via a microwave
field, which creates an unstable vacuum.

To create a metastable vacuum from this unstable vacuum, one can make
use of the classical concept of a modulated pendulum \citep{Stephenson1908,Kapitza1965_45,Kapitza1965_46,stoker1950nonlinear},
where a modulated amplitude of microwave coupling allows one to engineer
the metastable vacuum potential. We consider a symmetric intra-component
case in this work for its simplicity. In previous work, we suggested
the use of a pair of Zeeman states of $^{41}$K where $|1\rangle=|F=1,m_{F}=1\rangle$
and $|2\rangle=|F=1,m_{F}=0\rangle$ as the two spin states in the
proposed BEC experiment. This Zeeman pair has a symmetrical intra-component
Feshbach resonance where the inter-state scattering lengths between
the two components are zero \citep{Lysebo2010}.

We have shown in this paper that quantum vacuum nucleation is accelerated
at finite temperature, although the bubbles formed of true vacua may
be short-lived at high temperatures due to thermalization of the BEC.
The formation of the true vacuum bubbles at finite temperature generally
follows the expected behavior, apart from an accelerated tunneling
rate. Clearly, lower temperatures move one further into the true quantum
regime.

Our results also show that higher oscillator frequencies $\omega$
can remove short-wavelength instabilities, provided that there is
a high-momentum cutoff present. This increases the feasibility of
a false vacuum BEC experiment. The proposed table-top experiment using
Zeeman states of $^{41}$K may help to test vacuum tunneling theory
under real experimental conditions where thermal effects are unavoidable.
This provides insight into early universe models with high temperatures,
which is present in some early universe models \citep{zel1974cosmological}.

Such an experiment can detect a unique topological feature of the
scalar field vacuum. Unlike the von Neumann quantum entropy, which
is time-invariant, the topological phase entropy is predicted to reach
a maximum at the time when most tunneling occurs. Intriguingly, phase
entropy can decrease with time, because the final vacuum state is
more ordered than the state occurring while tunneling. This is a true
topological feature of the present model, since the relative phase
is only uniquely defined after a nonlocal unwrapping.

Such measurements may be useful in investigating proposals in which
a discretely broken symmetry provides a model for particle-antiparticle
symmetry. We find that domain wall formation is prevalent at high
temperatures, in agreement with qualitative predictions \citep{zel1974cosmological,kibble1980some}.
However, at low temperatures domain walls are restricted to universe
boundaries where they would be far removed from having a direct influence
on CMB inhomogeneities, which was thought to be a possible problem
with such theories.

Alternative implementations include a homogeneous, two dimensional
simulation. This allows even more complex topological vacuum structures
to form. Such experiments are realizable in micro-gravity with a shell
geometry \citep{garraway2016recent,lundblad2019shell}, for example
in the NASA CAL space-station environment. There are many other possible
approaches using different quantum technologies including superconducting
circuits or discrete Bose-Hubbard lattices, which are not analyzed
here. 
\begin{acknowledgments}
The authors acknowledge helpful discussions with Andrei Sidorov.
This research has been supported by the Australian Research Council
Discovery Project Grants schemes under Grants DP180102470 and DP190101480
.
\end{acknowledgments}

\section*{Appendix: Modulational instabilities}

Modulational instabilities can occur if the microwave modulation frequency
$\omega$ is too low relative to the momentum cutoff. In such a case,
the microwave modulation cannot be adiabatically eliminated, and parametric
instabilities occur. The effect of such Floquet modes was studied
by Braden et al. \citep{braden2018towards,Braden:2019aa}. In the
work above, we have included a momentum cutoff to prevent this, as
explained in the main text.

In this Appendix, we analyze the effects of modulational instabilities
at finite temperatures with a higher momentum cutoff. The value of
$\left\langle \cos\left(\phi_{a}\right)\right\rangle $ of the relative
phase of such a system evolves to around $\sim0$, indicating that
the transition from the false vacuum to a true vacuum is inhibited,
as the relative phase becomes completely randomized. This is studied
numerically by choosing a smaller lattice spacing and lartger $M$,
to resolve the high wave-number unstable modes.

The unstable modes occur in a narrow band centered at wave-numbers
as derived in \citep{braden2018towards,Braden:2019aa}. The dimensionless
critical unstable wavenumber is given by:

\begin{eqnarray}
\widetilde{k}_{c}^{2} & \approx & \frac{1}{2\widetilde{\nu}}\left(\sqrt{1+\widetilde{\omega}^{2}\widetilde{\nu}}-1\right)-\sigma,\label{eq:unstable_mode-1}
\end{eqnarray}
where $\sigma=\mathrm{cos}\phi_{a}=\pm1$ is the relative phase of
the fields.

\begin{figure}[h]
\begin{centering}
\includegraphics[width=0.6\columnwidth]{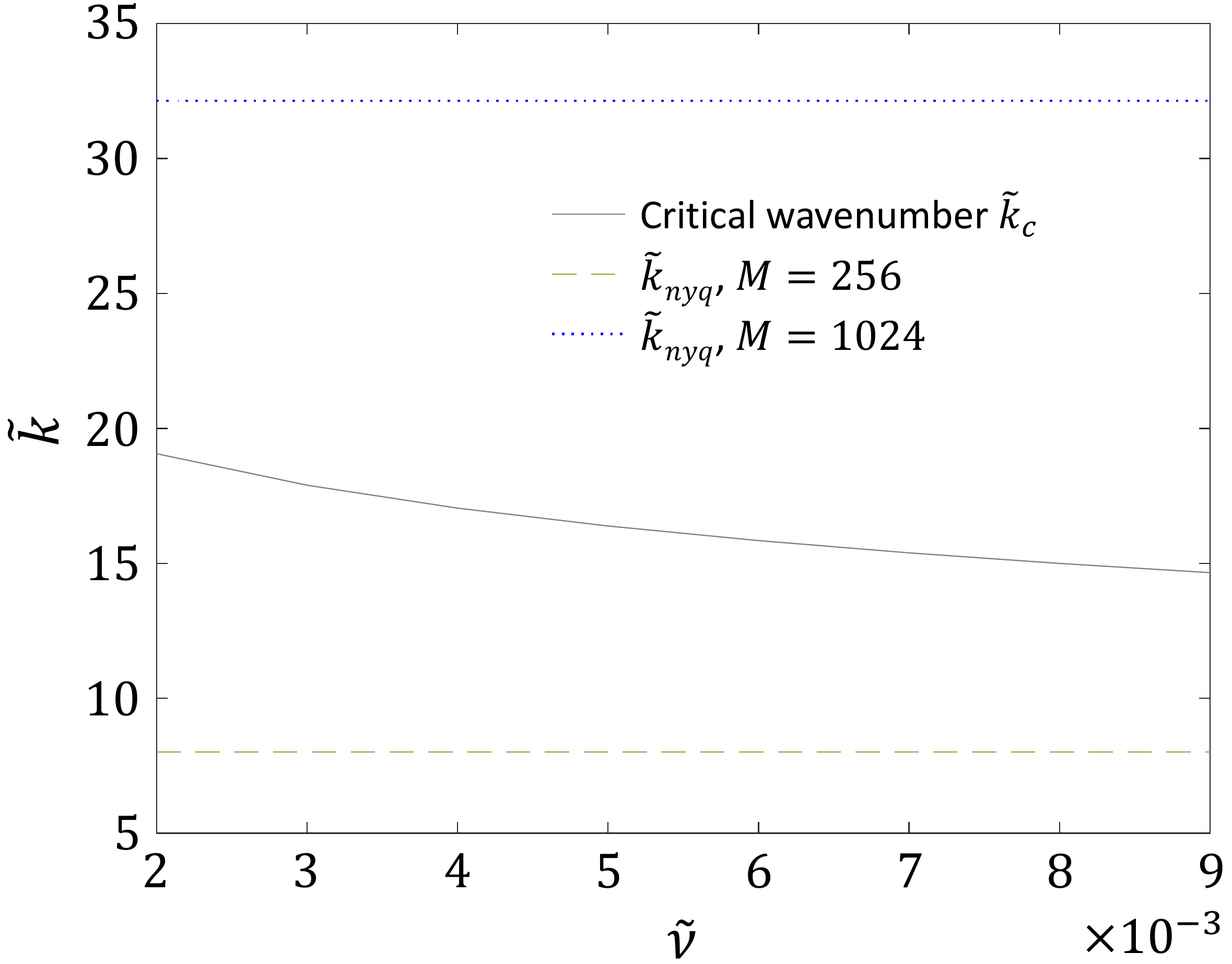}
\par\end{centering}
\caption{\label{fig:FVac unstable wavenumber-1}Critical wavenumber $\widetilde{k}_{c}$
and Nyquist wavenumber $\widetilde{k}_{nyq}=\pi/\Delta\widetilde{x}$
in the simulations for $\widetilde{\omega}=50$ at various values
of $\widetilde{\nu}$ with fixed length $\widetilde{L}=100$. For
$M=256$, $\widetilde{k}_{nyq}$ in the simulations are all below
$\widetilde{k}_{c}$, hence the unstable modes are excluded. For $M=1024$,
the unstable modes are included, as $\widetilde{k}_{nyq}>\widetilde{k}_{c}$.}
\end{figure}
Figure \ref{fig:FVac unstable wavenumber-1} shows the cutoff wave-numbers
$\widetilde{k}_{nyq}$ in the simulations of previous sections (\ref{subsec:Observational-criteria})
and (\ref{subsec:Thermally-induced-changes}) in comparison with the
corresponding critical wave-numbers $\widetilde{k}_{c}$ at various
values of $\widetilde{\nu}$. The Nyquist wavenumber determined by
the lattice spacing $\Delta\widetilde{x}$, is $\widetilde{k}_{nyq}=\pi/\Delta\widetilde{x}$.
The unstable wavenumber $\widetilde{k}_{c}$ is the wavenumber of
the highest gain Floquet mode.

As shown in Figure \ref{fig:FVac unstable wavenumber-1}, such effects
are excluded in the $M=256$ simulations presented in sections (\ref{subsec:Observational-criteria})
and (\ref{subsec:Thermally-induced-changes}) at all values of $\widetilde{\nu}$.
Therefore, the short-wavelength fluctuations between the true vacua
and the false vacua found at higher temperature (Figure \ref{fig:tau3e_4})
are purely due to the thermalization of the condensate. Thermal effects
were excluded in earlier work \citep{Braden:2019aa}. The dynamics
at finite temperatures presented in Figure \ref{fig:tau3e_4} shows
similarities to the dynamics in the presence of the Floquet modes
in \citep{Braden:2019aa}. In the following we will investigate the
combined effect of thermal fluctuations and Floquet instabilities
on the dynamics, by setting $\widetilde{k}_{nyq}$ above the critical
$\widetilde{k}_{c}$.

\begin{figure}[h]
\includegraphics[scale=0.48]{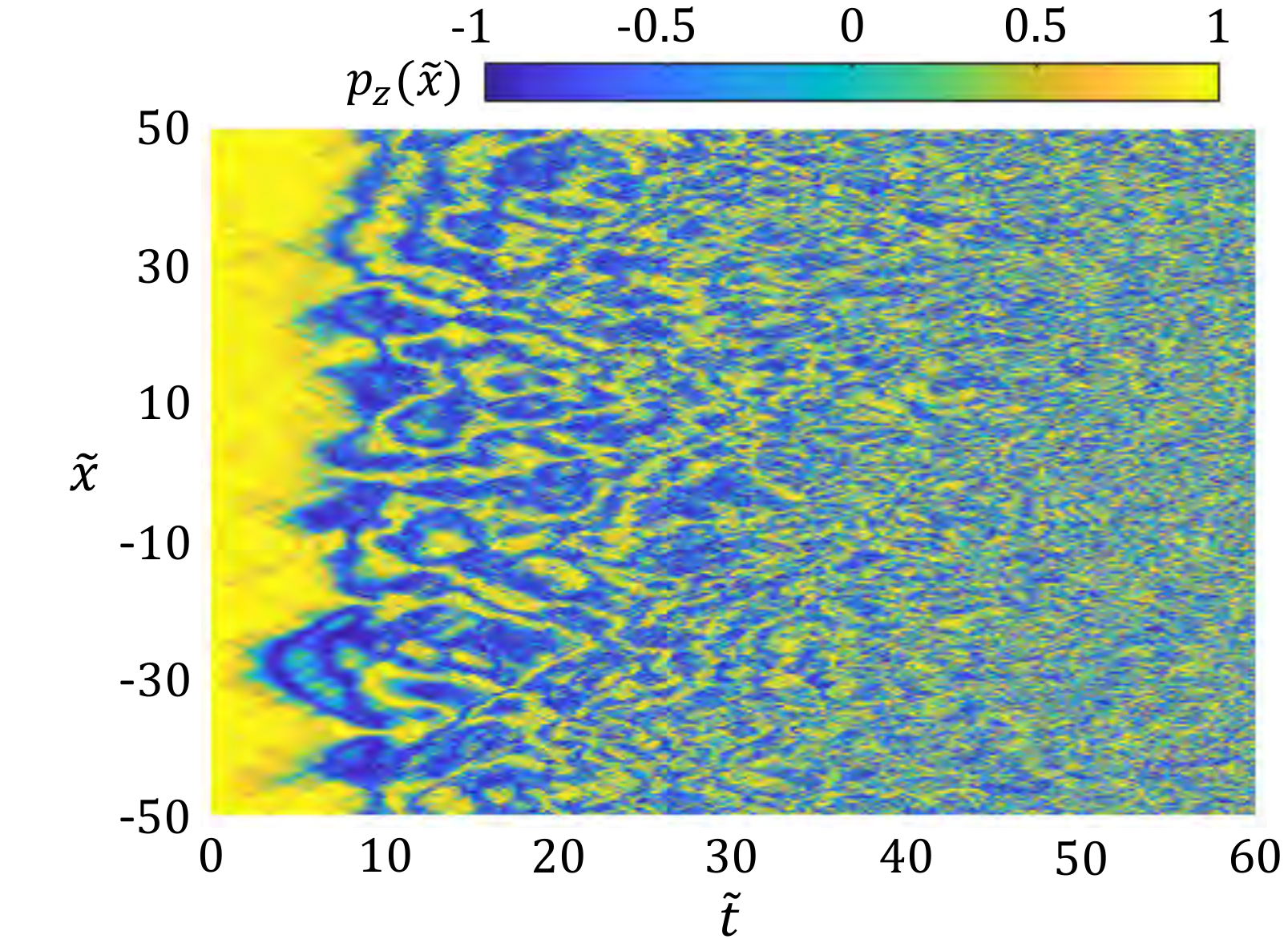}

\caption{\label{fig:Increase_M}Single-trajectory 1D false vacuum simulation
for the time evolution of $p_{z}$ at $\tau=1\times10^{-4}$ with
$M=1024$ to include the effect of Floquet modes. The modulation frequency
is $\widetilde{\omega}=50$. Dimensionless parameters are $\tilde{L}=100$,
$\lambda=1.2$, $\widetilde{\nu}=7\times10^{-3}$, $\widetilde{\rho}=200$.
Bubbles in the true vacua ($p_{z}=-1$) are short-lived and dominated
by fluctuations at later time.}
\end{figure}

Figure \ref{fig:Increase_M} shows our results when both thermal effects
and unstable Floquet modes are included. Here we reduce the lattice
spacing $\Delta\widetilde{x}$ in the simulations by increasing the
number of simulation modes to $M=1024$. This lattice spacing corresponds
to a Nyquist wavenumber of $\widetilde{k}_{nyq}\approx32.14$, well
above the critical wavenumber $\widetilde{k}_{c}\approx15.39$ for
$\widetilde{\nu}=7\times10^{-3}$.

The most significant effect of the unstable Floquet modes is that
true vacua formed at finite temperature are gradually destroyed, and
the system is eventually dominated by chaotic fluctuations. At a reduced
temperature of $\tau=1\times10^{-4}$, the presence of Floquet modes
causes fluctuations with wavelengths shorter than the dominant thermal
fluctuations. 

In Figure \ref{fig:tau1e_5}, \ref{fig:tau3e_4} and \ref{fig:tau1e_5_phase}
we show the effect of thermal fluctuations independently by setting
$\widetilde{k}_{nyq}<\widetilde{k}_{c}$ to exclude Floquet modes.
The true vacua in the absence of the Floquet modes have a metastable
structure with average relative phase $\langle\mathrm{cos}\phi_{a}\rangle$
acquiring a non-zero constant value (Figure \ref{fig:tau1e_5_phase}).
However, this behavior is different in the presence of the Floquet
modes, where the structure of the true vacua is short-lived. 

\begin{figure}[h]
\begin{centering}
\includegraphics[width=0.75\columnwidth]{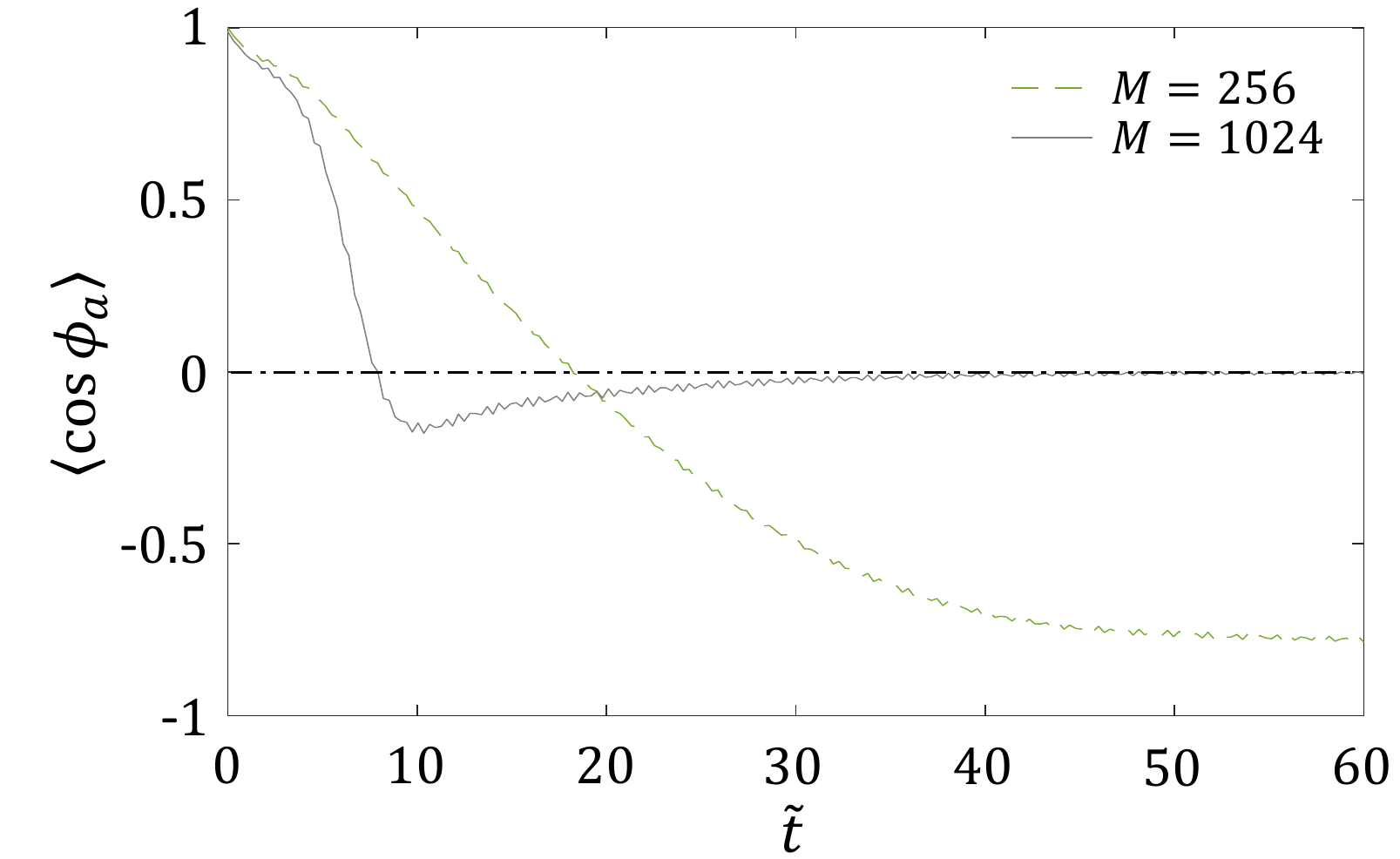}
\par\end{centering}
\caption{\label{fig:Increase_M_phase}Time evolution of average relative phase
$\langle\mathrm{cos}\phi_{a}\rangle$ using $8000$ trajectories for
$M=256$ ($\widetilde{k}_{nyq}\approx8.01<\widetilde{k}_{c}\approx15.39$,
Floquet modes excluded) and $M=1024$ ($\widetilde{k}_{nyq}\approx32.14>\widetilde{k}_{c}\approx15.39$,
Floquet modes included). All other parameters are as in Figure \ref{fig:Increase_M}.
The errors in $\langle\mathrm{cos}\phi_{a}\rangle$ are less than
$1\%$.}
\end{figure}

In the example presented in Figure \ref{fig:Increase_M}, most of
the vacuum bubbles only survive over a time duration $5\apprle\widetilde{t}\lesssim30$
(which corresponds to a real experimental time duration $\sim20\mathrm{ms}$).
At $\widetilde{t}\apprge30$, one can expect the average relative
phase $\langle\mathrm{cos}\phi_{a}\rangle$ to be around $\sim0$
as the chaotic fluctuations dominate. This is confirmed by the averaged
result using $8000$ trajectories shown in Figure \ref{fig:Increase_M_phase}.

\begin{figure}[h]
\begin{centering}
\includegraphics[width=0.75\columnwidth]{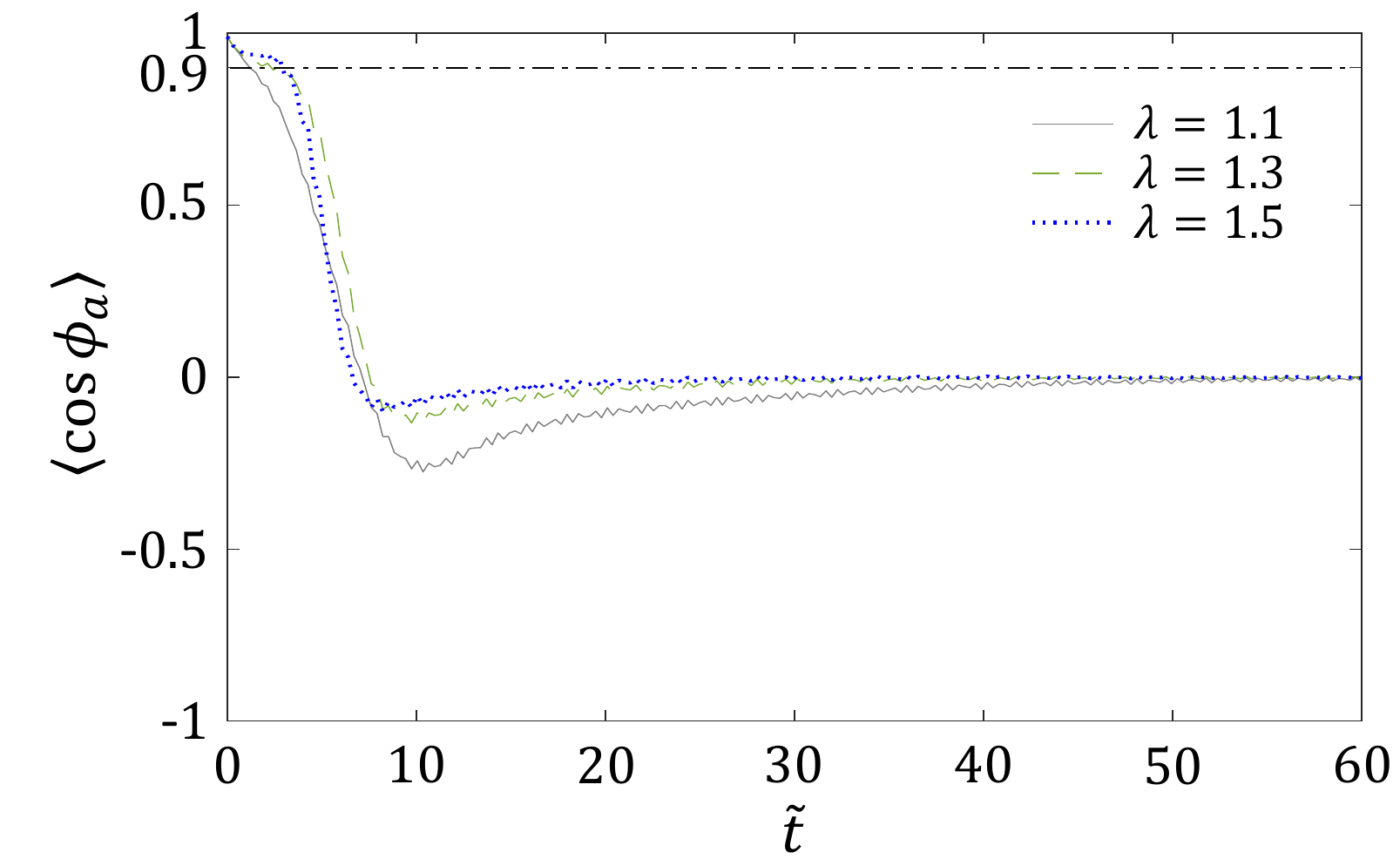}
\par\end{centering}
\caption{\label{fig:lambda}Time evolution of average relative phase $\langle\mathrm{cos}\phi_{a}\rangle$
using $8000$ trajectories for $\lambda=1.1$ (solid line), $\lambda=1.3$
(dash line), and $\lambda=1.5$ (dot line). Floquet modes are included
by setting number of modes $M=1024$ ($\widetilde{k}_{nyq}\approx32.14>\widetilde{k}_{c}\approx15.39$).
Reduced temperature $\tau=1\times10^{-4}$, other dimensionless parameters
$\widetilde{L}=100$, $\widetilde{\omega}=50$, $\widetilde{\nu}=7\times10^{-3}$,
$\widetilde{\rho}=200$. The errors in $\langle\mathrm{cos}\phi_{a}\rangle$
are less than $1\%$.}
\end{figure}

The modulation depth $\lambda$ is known to correspond to the strength
of the unstable Floquet modes \citep{braden2018towards}. In the absence
of thermal effects, earlier studies \citep{Braden:2019aa} showed
that increasing the modulation depth $\lambda$ can reduce the time-scale
of the Floquet modes, and results in the stabilization of long-wavelength
structure in the decay of false vacuum. This effectively delays the
nucleation of the true vacuum bubbles. For systems at finite temperature
where short-wavelength thermal fluctuations coexist with the true
vacua, we found that this delay of bubble nucleation is valid. Figure
\ref{fig:lambda} shows the examples on the effect of increasing $\lambda$
at a fixed reduced temperature $\tau=1\times10^{-4}$.

At time $\widetilde{t}\lesssim10$, our results show that the systems
with larger $\lambda$ ($\lambda=1.3,1.5$) break through the threshold
value of the tunneling initiation $\langle\mathrm{cos}\phi_{a}\rangle=0.9$
later than the $\lambda=1.1$ system, indicating a delay of the bubble
nucleation. In the presence of Floquet instabilities, increasing $\lambda$
results in a significant reduction of the average relative phase.
The peak value of $|\langle\mathrm{cos}\phi_{a}\rangle|$ drops from
$\sim0.35$ to $\sim0.1$ as $\lambda$ is increased from $1.1$ to
$1.5$. For decay at finite temperature, we have shown in the main
text Figure \ref{fig:tau1e_5_phase} that this phase signal is also
weakened by the influence of thermal fluctuations in the absence of
the Floquet instabilities.

Increasing $\lambda$ in the presence of Floquet modes further reduces
the phase signal, which increases the difficulty of measuring true
vacua in an experiment. Therefore, it is expected that true vacuum
bubbles are destroyed by Floquet instabilities regardless of $\lambda$.
At time $\widetilde{t}\gtrsim10$ in Figure \ref{fig:lambda}, the
average relative phase of all three systems reaches $\sim0$ eventually.
The true vacuum bubbles in the system with the lowest $\lambda$ survive
longer due to their larger sizes. In the case where Floquet instabilities
are not negligible, the tuning of the modulation depth $\lambda$
plays a role in the balance between time duration and strength of
true vacua signals.

\begin{figure}[h]
\includegraphics[scale=0.48]{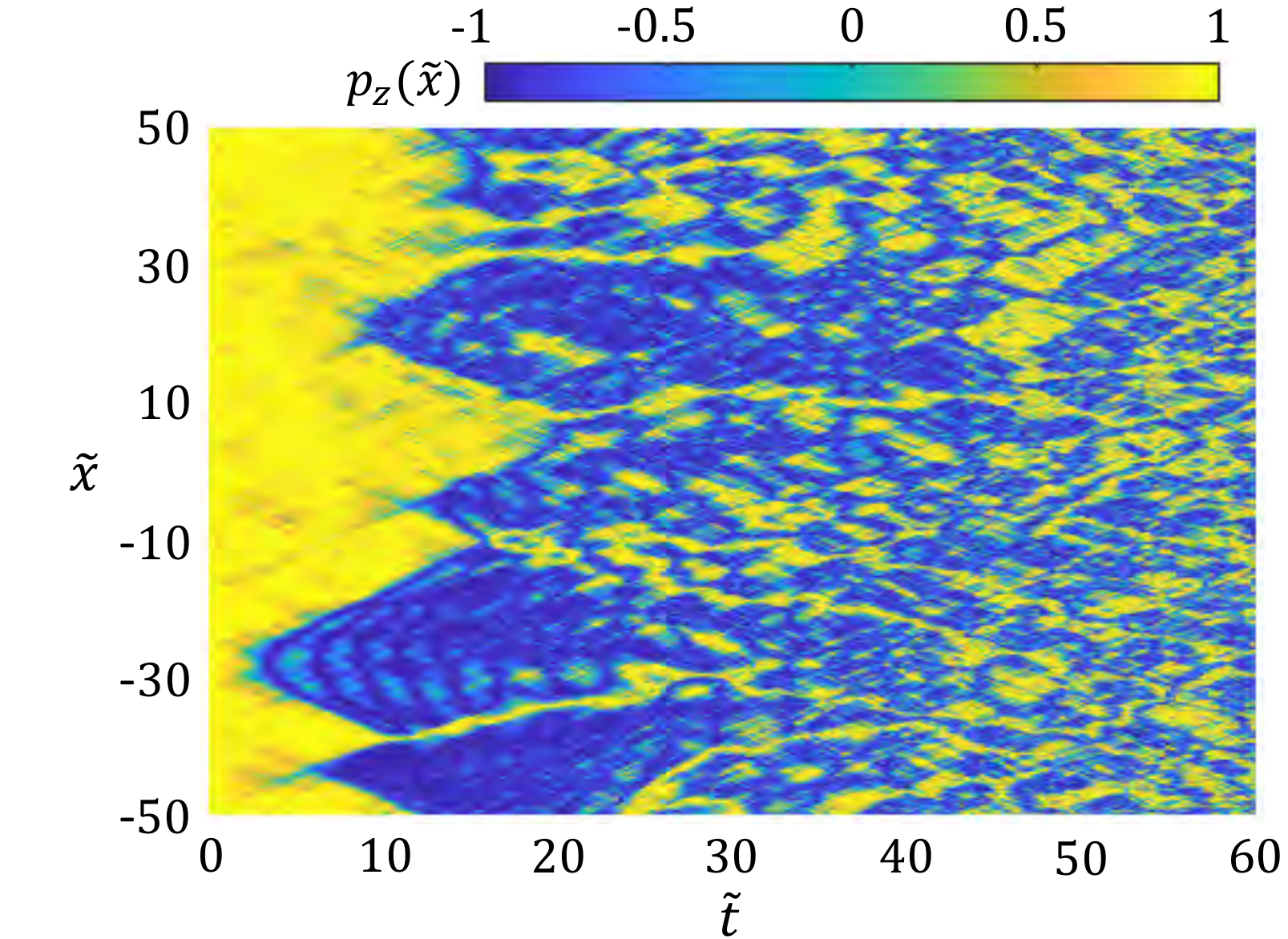}

\caption{\label{fig:omega}Single-trajectory simulation for $p_{z}$ at $\tau=1\times10^{-4}$
with $M=1024$. The modulation frequency is $\widetilde{\omega}=150$.
Floquet modes are included: $\widetilde{k}_{nyq}\approx32.1>\widetilde{k}_{c}\approx28.8$.
Dimensionless parameters $\tilde{L}=100$, $\lambda=1.2$, $\widetilde{\nu}=7\times10^{-3}$,
$\widetilde{\rho}=200$. Comparing to Figure \ref{fig:Increase_M},
the true vacuum ($p_{z}=-1$) is relatively long-lived and stable
.}
\end{figure}

In order to remove the unstable Floquet modes from the system, one
can increase the oscillation frequency $\widetilde{\omega}$ to shift
$\widetilde{k}_{c}$ to higher wave-numbers above the increased momentum
cutoff used in this Appendix. Figure \ref{fig:omega} illustrates
the simulated dynamics of the false vacuum at a fixed temperature
$\tau=1\times10^{-4}$ , and increasing $\widetilde{\omega}$.

In Figure \ref{fig:omega}, the results show that the stabilization
of true vacua is sensitive to an increase of $\widetilde{\omega}$.
Both the lifetime and the phase signal of the true vacua are enhanced.
In comparison with Figure \ref{fig:Increase_M} ($\widetilde{\omega}=50$
at $\tau=1\times10^{-4}$), increasing $\widetilde{\omega}$ to $150$
(Figure \ref{fig:omega}) extends the sizes of the true vacuum bubbles
and the chaotic fluctuations are suppressed in the true vacua. The
survival time of the true vacuum bubbles before domination by fluctuations
is also improved slightly when $\widetilde{\omega}$ is increased,
from roughly $\widetilde{t}\lesssim30$ for $\widetilde{\omega}=50$
to $\widetilde{t}\apprle40$ for $\widetilde{\omega}=150$.

As mentioned, the simulated system with $\widetilde{\omega}=50$ in
Figure \ref{fig:Increase_M} includes the Floquet modes by setting
$\widetilde{k}_{nyq}\approx32.14$ well above the critical wavenumber
$\widetilde{k}_{c}\approx15.39$; while for the system with $\widetilde{\omega}=150$,
the Floquet modes are partially removed as the critical wavenumber
is increased to $\widetilde{k}_{c}\approx28.79$. This partial removal
of the Floquet modes reduces the chaotic fluctuations and results
in the partial stabilization of true vacua.

\begin{figure}[h]
\includegraphics[scale=0.48]{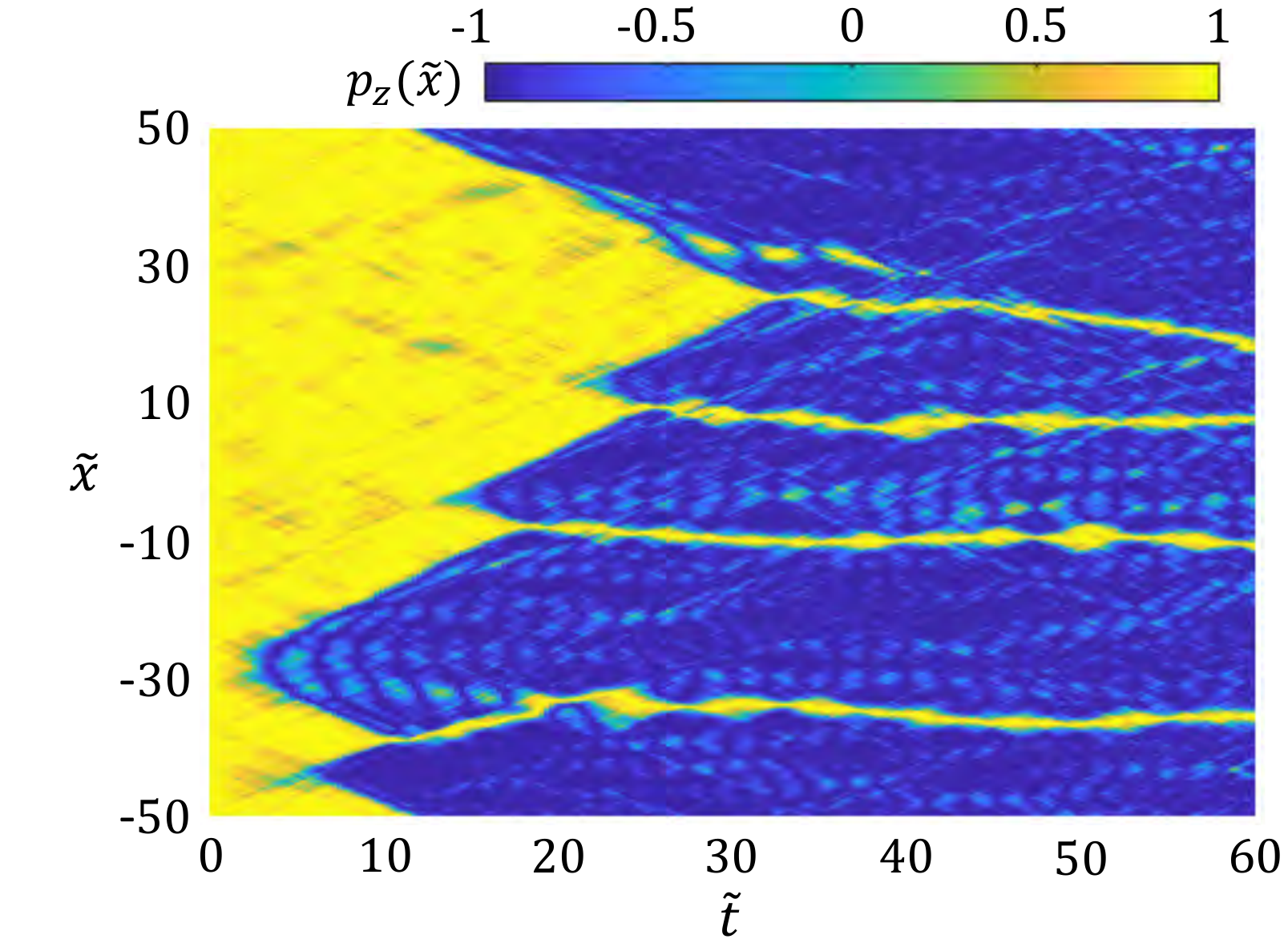}

\caption{\label{fig:omega_200}Single-trajectory simulation for the time evolution
of $p_{z}$ with an increased frequency $\widetilde{\omega}=200$
to remove the Floquet modes ($\widetilde{k}_{nyq}\approx32.14<\widetilde{k}_{c}\approx33.57$).
Other parameters are as in Figure \ref{fig:omega}.}
\end{figure}

If we further increase the modulation frequency to $\widetilde{\omega}=200$,
the critical wavenumber is shifted to $\widetilde{k}_{c}\approx33.57$
which is slightly higher than $\widetilde{k}_{nyq}$. In this case,
the Floquet modes are almost completely excluded. Figure \ref{fig:omega_200}
shows that in the dynamics of the false vacuum with $\widetilde{\omega}=200$,
one can see that the size of the true vacuum bubbles are extended,
and a break-down of the vacuum bubbles is not observed in the simulation.

\begin{figure}[h]
\includegraphics[width=0.75\columnwidth]{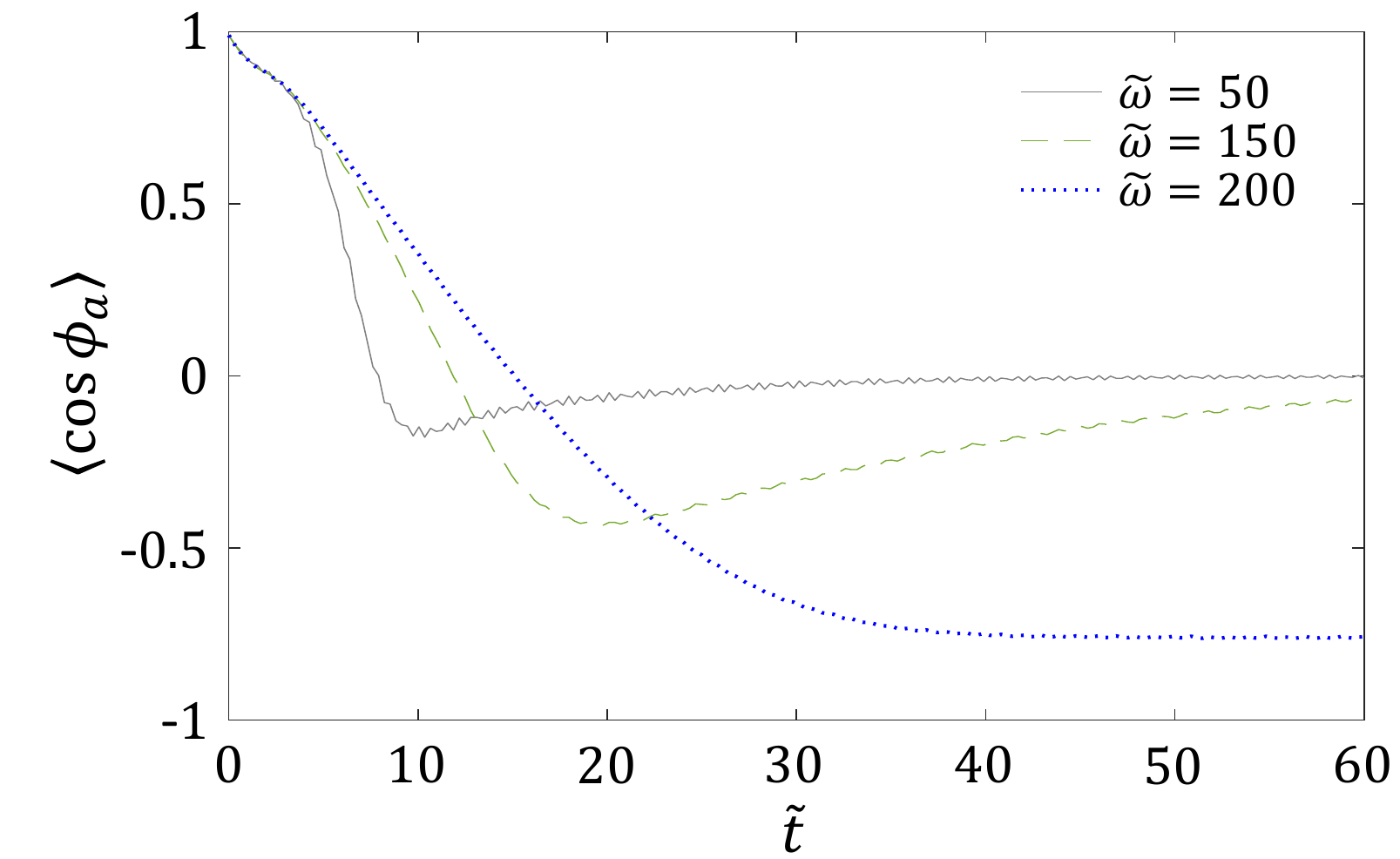}

\caption{\label{fig:omega_phase}Time evolution of $\langle\mathrm{cos}\phi_{a}\rangle$
using $8000$ trajectories for $\widetilde{\omega}=50$ (solid line),
$150$ (dash line) and $200$ (dot line), all other parameters are
as in Figure \ref{fig:omega}. The errors in $\langle\mathrm{cos}\phi_{a}\rangle$
are less than $1\%$.}
\end{figure}

Increasing $\widetilde{\omega}$ to suppress the chaotic fluctuations
induced by the presence of the Floquet instabilities has a significant
effect on the stabilization of true vacua. Figure \ref{fig:omega_phase}
shows that the suppression of the chaotic fluctuations enhances the
average relative phase $\langle\mathrm{cos}\phi_{a}\rangle$. The
relative phase can reach $\langle\mathrm{cos}\phi_{a}\rangle\approx-0.75$
even at the high reduced temperature $\tau=1\times10^{-4}$, which
is comparable to a measurement at a lower temperature $\tau=1\times10^{-5}$
($\langle\mathrm{cos}\phi_{a}\rangle\approx-0.8\sim-0.9$, in Figure
\ref{fig:tau1e_5_phase}).

In summary, bubble nucleation and break-down are delayed by increasing
$\widetilde{\omega}$ to remove the Floquet modes. From the averaged
results using $8000$ trajectories, true vacua reach their peak size
at a time $\widetilde{t}\approx10$ for $\widetilde{\omega}=50$ and
$\widetilde{t}\approx20$ for $\widetilde{\omega}=150$, and are later
destroyed by chaotic fluctuations due to Floquet modes (i.e. $\langle\mathrm{cos}\phi_{a}\rangle\approx0$).
For $\widetilde{\omega}=200$, the time of the peak relative phase
is delayed to $\widetilde{t}\approx35$ and $\langle\mathrm{cos}\phi_{a}\rangle$
remains $\sim-0.8$ over the remaining simulation time duration. The
high dimensionless frequency $\widetilde{\omega}=200$ in the simulations
corresponds to an oscillation frequency $\omega=2\pi\times38.24\mathrm{kHz}$,
which is achievable in current experiments. We emphasize that this
still requires an overall momentum cutoff, although at a higher wavenumber.

\bibliographystyle{apsrev4-1}

\end{document}